\documentclass[aps,pra,onecolumn,showpacs,notitlepage,superscriptaddress,letterpaper]{revtex4-1}
\usepackage{graphicx}
\usepackage{amsmath}
\usepackage{esint}
\usepackage{verbatim}
\usepackage{color}
\usepackage{SIunits}
\usepackage[colorlinks, linkcolor=blue, citecolor=blue, urlcolor=blue,breaklinks=true]{hyperref}
\newcommand{\Tf}{T_{f}}
\newcommand{\nfm}{n_{f,m}}
\newcommand{\Tm}{T_{m}}

\newcommand{\SiN}{\text{Si}_{3}\text{N}_4}
\newcommand{\GammaAS}{\Gamma_{\text{AS}}}

\newcommand{\kappai}{\kappa_{\text{i}}}
\newcommand{\kappae}{\kappa_{\text{e}}}

\newcommand{\gammai}{\gamma_{m,i}}% intrinsic mechanical (energy) damping rate
\newcommand{\gammam}{\gamma_{m}}% total mechanical (energy) damping rate
\newcommand{\omegam}{\omega_{m}}

\newcommand{\xzpf}{x_{\text{zpf}}}% mechanical zpf amplitude
\newcommand{\meff}{m_{\text{eff}}}% mechanical motional mass
\newcommand{\bhat}{\hat{b}}
\newcommand{\bdag}{\hat{b}^{\dagger}}

\newcommand{\ahat}{\hat{a}}
\newcommand{\adag}{\hat{a}^{\dagger}}
\newcommand{\ncavp}{n_p}% cavity occupancy due to probe
\newcommand{\omegap}{\omega_{p}}% microwave probe frequency
\newcommand{\omegad}{\omega_{d}}% microwave drive frequency
\newcommand{\ncavd}{n_{d}}% cavity occupancy due to drive
\newcommand{\Deltard}{\Delta_{r,d}}% resonator-drive detuning
\newcommand{\gzeroE}{g_{\text{0}}}% vacuum coupling between circuit and mechanical resonator
\newcommand{\kappaE}{\kappa}% microwave cavity total energy damping rate 
\newcommand{\kappaEe}{\kappa_{e}}% microwave cavity external energy coupling rate
\newcommand{\kappaEi}{\kappa_{i}}% microwave cavity intrinsic energy damping rate 
\newcommand{\omegacE}{\omega_{r}}% microwave cavity resonance frequency
\newcommand{\gammaOME}{\gamma_{\text{EM}}}
\newcommand{\GOM}{G}
\newcommand{\HOM}{H_{\text{OM}}}

\newcommand{\Ctot}{C_{\text{tot}}}
\newcommand{\Cm}{C_{m}}
\newcommand{\Cs}{C_{s}}
\newcommand{\Cl}{C_{l}}
\newcommand{\gEM}{g_{\text{EM}}}
\newcommand{\omegasrf}{\omega_{\text{coil}}}
\newcommand{\Acoil}{A_{\text{coil}}}
\newcommand{\lbeam}{l_{b}}
\newcommand{\Wx}{W_{x}}
\newcommand{\Wy}{W_{y}}
\newcommand{\WAl}{W_{\text{Al}}}
\newcommand{\tmem}{t_{\text{mem}}}
\newcommand{\tAl}{t_{\text{Al}}}
\newcommand{\Zo}{Z_{0}}

\newcommand{\Pd}{P_{d}}
\newcommand{\Pref}{P_{\text{ref}}}
\newcommand{\OmegaR}{\Omega_{R}}% bare Rabi frequency of TLS-resonator due to microwave drive
\newcommand{\gammaTLS}{\gamma_\text{tls}}
\newcommand{\gzeroTLS}{g_\text{0,tls}}
\newcommand{\omegaTLS}{\omega_\text{tls}}
\newcommand{\omegatildeTLS}{\widetilde{\omega}_\text{tls}}

\newcommand{\DeltaTLSd}{\Delta_\text{tls,d}}

\newcommand{\dipoleTLS}{\vec{d}}
\newcommand{\nwb}{n_{b,\text{wg}}}
\newcommand{\nrb}{n_{b,r}}
\newcommand{\nmb}{n_{b,m}}
\newcommand{\nr}{n_{r}}
\newcommand{\nadd}{n_{\text{add}}}
\newcommand{\nbar}{n_{m}}
\newcommand{\aop}{\hat{a}}
\newcommand{\bop}{\hat{b}}

\newcommand{\ain}{\hat{a}_\mathrm{in}}
\newcommand{\aout}{\hat{a}_\mathrm{out}}
\newcommand{\ab}{\hat{a}_{b,r}}
\newcommand{\awb}{\hat{a}_{b,\text{wg}}}
\newcommand{\bb}{\hat{b}_{b,m}}

\newcommand{\chir}{\chi_\mathrm{r}}
\newcommand{\chirt}{\widetilde{\chi}_\mathrm{r}}
\newcommand{\chim}{\chi_\mathrm{m}}
\newcommand{\chimt}{\widetilde{\chi}_\mathrm{m}}

\begin{document}
%title / authors / affils
\title{Quantum Electromechanics on Silicon Nitride Nanomembranes}
\author{J.~M.~Fink}
%\email{jfink@caltech.edu}
\altaffiliation[Current address: ]{Institute of Science and Technology Austria (IST Austria), 3400 Klosterneuburg, Austria}
\affiliation{Kavli Nanoscience Institute and Thomas J. Watson, Sr., Laboratory of Applied Physics, California Institute of Technology, Pasadena, CA 91125, USA}
\affiliation{Institute for Quantum Information and Matter, California Institute of Technology, Pasadena, CA 91125, USA}
\author{M.~Kalaee}
\affiliation{Kavli Nanoscience Institute and Thomas J. Watson, Sr., Laboratory of Applied Physics, California Institute of Technology, Pasadena, CA 91125, USA}
\affiliation{Institute for Quantum Information and Matter, California Institute of Technology, Pasadena, CA 91125, USA}
\author{A.~Pitanti}
\altaffiliation[Current address: ]{NEST, Istituto Nanoscienze-CNR and Scuola Normale Superiore, I-56126 Pisa, Italy}
\affiliation{Kavli Nanoscience Institute and Thomas J. Watson, Sr., Laboratory of Applied Physics, California Institute of Technology, Pasadena, CA 91125, USA}
\affiliation{Institute for Quantum Information and Matter, California Institute of Technology, Pasadena, CA 91125, USA}
\author{R.~Norte}
\altaffiliation[Current address: ]{Kavli Institute of Nanoscience, Delft University of Technology, 2600 GA, Delft, The Netherlands}
\affiliation{Kavli Nanoscience Institute and Thomas J. Watson, Sr., Laboratory of Applied Physics, California Institute of Technology, Pasadena, CA 91125, USA}
\affiliation{Institute for Quantum Information and Matter, California Institute of Technology, Pasadena, CA 91125, USA}
\author{L.~Heinzle}
%\affiliation{Kavli Nanoscience Institute and Thomas J. Watson, Sr., Laboratory of Applied Physics, California Institute of Technology, Pasadena, CA 91125, USA}
%\affiliation{Institute for Quantum Information and Matter, California Institute of Technology, Pasadena, CA 91125, USA}
\affiliation{Department of Physics, ETH Z\"urich, CH-8093 Z\"urich, Switzerland}
\author{M.~Davan\c{c}o}
\affiliation{Center for Nanoscale Science and Technology, National Institute of Standards and Technology, Gaithersburg, MD 20899, USA}
\author{K.~Srinivasan}
\affiliation{Center for Nanoscale Science and Technology, National Institute of Standards and Technology, Gaithersburg, MD 20899, USA}
\author{O.~Painter}
\email{opainter@caltech.edu}
\affiliation{Kavli Nanoscience Institute and Thomas J. Watson, Sr., Laboratory of Applied Physics, California Institute of Technology, Pasadena, CA 91125, USA}
\affiliation{Institute for Quantum Information and Matter, California Institute of Technology, Pasadena, CA 91125, USA}

\date{\today}

% ABSTRACT - 150 words - unreferenced
\begin{abstract}
We present a platform based upon silicon nitride nanomembranes for integrating superconducting microwave circuits with planar acoustic and optical devices such as phononic and photonic crystals.  Utilizing tensile stress and lithographic patterning of a silicon nitride nanomembrane we are able to reliably realize planar capacitors with vacuum gap sizes down to $s \approx 80$~nm.  In combination with spiral inductor coils of micron pitch, this yields microwave ($\approx 8$~GHz) resonant circuits of high impedance ($Z_{0} \approx 3.4$~k$\Omega$) suitable for efficient electromechanical coupling to nanoscale acoustic structures.  We measure an electromechanical vacuum coupling rate of $\gzeroE/2\pi = 41.5$~Hz to the low frequency ($4.48$~MHz) global beam motion of a patterned phononic crystal nanobeam, and through parametric microwave driving reach a backaction cooled mechanical mode occupancy as low as $\nbar = 0.58$.  
\end{abstract}

% \pacs{}
\maketitle
%
%%%INTRODUCTION up to 500 words

Thin films of silicon nitride ($\SiN$), when grown stoichiometrically via low-pressure chemical vapor deposition (LPCVD) on silicon substrates, can be used to form membranes with large tensile stress ($\approx 1$~GPa), thickness down to tens of nanometers and planar dimensions as large as centimeters~\cite{NORCADA}.  The large tensile stress of these films allows one to pattern membranes into extreme aspect ratio nanostructures which maintain precise planarity and alignment~\cite{Cohen2013,Yu2014b}.  The additional energy stored in tension also results in a significant reduction in mechanical damping in high tension nitride films~\cite{Liu2009b,Unterreithmeier2010,Yu2012,Norte2015}, with $Q$-frequency products as large as $2\times 10^{13}$~Hz and $3 \times 10^{15}$~Hz having been observed at room temperature~\cite{Wilson2009} and milliKelvin temperatures~\cite{Yuan2015b}, respectively.  As an optical material, $\SiN$ thin films have been used to support low loss guided modes for microphotonic applications, with a measured loss tangent in the near-IR of $< 3 \times 10^{-7}$~\cite{Barclay2006}.

Owing to their unique elastic and dielectric properties, $\SiN$ nanomembranes have recently been utilized in a variety of cavity-optomechanical and cavity-electromechanical experiments~\cite{Aspelmeyer2014} involving the interaction of membrane motion and radiation pressure of either optical or microwave light.  These experiments include optical back-action cooling of a millimeter-scale membrane close to its quantum ground state of motion~\cite{Thompson2008,Wilson2009,Peterson2015,Underwood2015}, measurement of radiation pressure shot noise~\cite{Purdy2013} and optical squeezing~\cite{Purdy2013b}, and parametric conversion between optical and microwave photons~\cite{Andrews2014}.  Thin film $\SiN$ has also been patterned into various other optomechanical geometries, such as deformable photonic crystals~\cite{Eichenfield2009a}, nanobeams coupled to microdisk resonators~\cite{Anetsberger2009}, and optomechanical crystal cavities which can be used to co-localize (near-IR) photons and (GHz) phonons into wavelength-scale modal volumes~\cite{Eichenfield2009,Grutter2015}.      

Here we explore $\SiN$ nanomembranes as a low-loss substrate for integrating superconducting microwave circuits and planar nanomechanical structures.  In particular, we exploit the thinness of the nanomembrane to reduce parasitic capacitance and greatly increase the attainable impedance of the microwave circuit.  We also use the in-plane stress to engineer the post-release geometry of a patterned membrane~\cite{Camacho2009,Grutter2015}, resulting in planar capacitors with vacuum gaps down to tens of nanometers.  Combining the large capacitance of planar vacuum gap capacitors and the low stray capacitance of compact spiral inductor coils formed on a $\SiN$ nanomembrane, we show theoretically that it is possible to realize large electromechanical coupling to both in-plane flexural modes and localized phononic bandgap modes of a patterned beam structure.  Two-tone microwave measurements of an $8$~GHz $LC$ circuit at milliKelvin temperatures in a dilution refrigerator confirm the predictions of strong electromechanical coupling to the low-frequency flexural mode of such a beam, and microwave backaction damping is used to cool the mechanical resonance to an average phonon occupancy of $\nbar = 0.58$.  These results, along with recent theoretical and experimental efforts to realize $\SiN$ optomechanical crystals~\cite{Davanco2012b,Grutter2015}, indicate the viability of $\SiN$ nanomembranes as an all-in-one substrate for quantum electro-opto-mechanical experiments.  Such membrane systems could be used, for instance, to realize a chip-scale quantum optical interface to superconducting quantum circuits~\cite{Regal2011,Safavi2011,Barzanjeh2011,Bochmann2013,Andrews2014,Bagci2014}. 

\begin{figure*}[tp!]
\begin{center}
\includegraphics[width=1.0\textwidth]{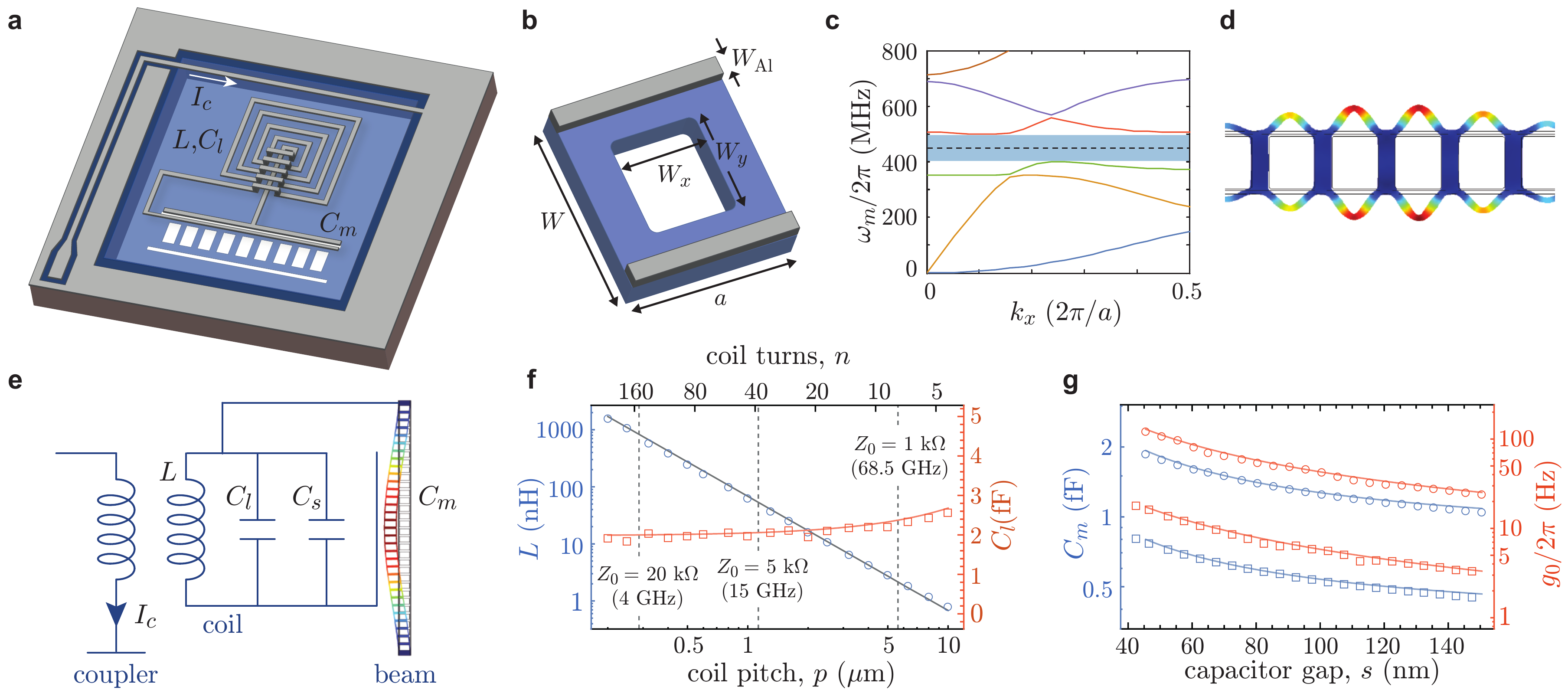}
\caption{\textbf{Device design.} 
\textbf{a,} Schematic of the membrane electromechanical circuit.  
\textbf{b}, Unit cell of the phononic crystal nanobeam. 
\textbf{c}, Acoustic band diagram of the phononic crystal nanobeam with $a=W=2.23$~$\mu$m, $\Wx=\Wy=1.52$~$\mu$m, and $\WAl=170$~nm.  The nitride membrane thickness and aluminum wire thickness are $\tmem=300$~nm and $\tAl=65$~nm, respectively.  The acoustic bandgap is shaded in blue, with the localized breathing mode frequency indicated as a dashed line.
 \textbf{d}, Plot of the FEM-simulated breathing mode profile.  Mechanical motion is indicated by an exaggerated displacement of the beam structure and by color, with red (blue) color indicating regions of large (small) amplitude of the motion.
\textbf{e}, Electrical circuit diagram, where $I_\mathrm{c}$ is the current through the reflective coupler, $L$ is the coil inductance, $\Cl$ is the coil capacitance, $\Cs$ is additional stray capacitance, and $\Cm$ is the motional capacitance. The simulated displacement of the in-plane fundamental flexural mode of the beam is shown.
\textbf{f}, Inductance ($L$) and capacitance ($\Cl$) of a planar square coil inductor of constant area $\Acoil = 87$~$\mu$m $\times 87$~$\mu$m and variable wire-to-wire pitch $p$.  Wire width and thickness are $500$~nm and $120$~nm, respectively.  Method of moments~\cite{SONNET} numerically simulated values are shown as open circles (inductance) and open squares (capacitance).  Calculations using an analytical model of the planar coil inductor~\cite{Mohan1999} are shown as a solid line.  Vertical lines are shown for coils with a characteristic impedance of $\Zo = 1$~k$\Omega$, $5$~k$\Omega$, and $20$~k$\Omega$, with the coil self resonance frequency indicated in brackets.
\textbf{g}, FEM simulations of the modulated capacitance $\Cm$ (blue symbols) and the electromechanical coupling $\gzeroE/2\pi$ (red symbols) of the in-plane fundamental flexural mode (circles) and the phononic crystal breathing mode (squares) as a function of the capacitor gap size $s$.  Solid curves indicate a $1/s$ fit to the capacitance and coupling data.} 
\label{fig:model}
\end{center}
\end{figure*}

\section*{Device Design and Fabrication}
\label{sec:design_fab}

The key elements of the membrane microwave circuits studied in this work are shown schematically in Fig.~\ref{fig:model}(a).  The circuits are created through a series of patterning steps of an aluminum-coated $300$~nm thick $\SiN$ nanomembrane, and consist of a mechanical beam resonator, a planar vacuum gap capacitor, a spiral inductor ($L$), and a $50$~Ohm coplanar waveguide feedline.  The vacuum gap capacitor, formed across the nanoscale cuts in the membrane defining the beam resonator, is connected in parallel with the coil inductor to create an $LC$ resonator in the microwave C band.  Each $LC$ resonator sits within a $777$~$\mu$m $\times 777$~$\mu$m square membrane and is surrounded on all sides by a ground plane.  The coplanar waveguide feedline is terminated by extending the center conductor from one side of the membrane to the other, where it is shorted to the ground plane.  Electrical excitation and read-out of the $LC$ resonator is provided by inductive coupling between the center conductor and the spiral inductor.  Note that although thinner membranes could have been used, our choice of a $300$~nm thick membrane allows for compatibility with single-mode near-IR photonic devices, and is guided by an ultimate goal of integrating planar optical components with electromechanical ones as per Ref.~\cite{Davanco2012b}.    
 
%Figure~\ref{fig:fab} shows the details of our fabrication procedure.  

%Thinner nitride layers could also be used, however a $300$~nm thick layer was chosen as a compromise in terms of reducing parasitic capacitance (and loss) for the microwave circuit and being compatible with guided wave optics in the near-IR.

The electromechanical coupling between the beam resonator and the $LC$ circuit in general depends upon the particular resonant mode of the beam, and is given in terms of the linear dispersion ($\gEM$) of the microwave circuit resonance frequency ($\omegacE$) with respect to modal amplitude coordinate $u$,

\begin{equation}
\gEM = \frac{\partial\omegacE}{\partial u}=-\eta \frac{\omegacE}{2C_\mathrm{m}} \frac{\partial C_\mathrm{m}}{\partial u}.
\label{eq:coupling}
\end{equation}

\noindent Here $\Cm$ is the vacuum gap capacitance across the beam, $\Ctot$ is the total capacitance of the circuit, and $\eta\equiv \Cm/\Ctot$ is the motional participation ratio.  In the case of uniform in-plane beam motion, and assuming $\Cm$ behaves approximately as a parallel plate capacitor, the cavity dispersion simplifies to $\gEM = \eta \left(\omegacE/2s_{0}\right)$, where $s_{0}$ is the nominal capacitor gap size.  The vacuum coupling rate, describing the interaction between light and mechanics at the quantum level, is given by $\gzeroE \equiv \gEM \xzpf$, where $\xzpf = (\hbar/2\meff\omegam)^{1/2}$ is the zero-point amplitude, $\meff$ is the motional mass, and $\omegam$ is the mechanical resonance frequency of a given mechanical mode of the beam.

In this work we consider a patterned beam resonator of width $W=2.23$~$\mu$m and length $\lbeam=71.4$~$\mu$m which supports two in-plane resonant modes which can be coupled efficiently to microwave or optical cavities~\cite{Davanco2012b}. The beam unit cell, shown in Fig.~\ref{fig:model}(b), has a lattice constant $a$ and contains a central hole of width $\Wx$ and height $\Wy$.  A pair of upper and lower aluminum wires of thickness $65$~nm and width $170$~nm at the edges of the beam form one half of the vacuum gap capacitor electrodes.  Simulations of the mechanical modes of the beam are performed using a finite-element method (FEM) solver~\cite{COMSOL}, and include the internal stress of the nitride film ($\sigma \approx 1$~GPa).

The simulated fundamental in-plane flexural mode of the patterned and wired beam, a displacement plot of which is inserted into the microwave circuit of Fig.~\ref{fig:model}(e), occurs at a frequency of $\omegam/2\pi = 4.18$~MHz.  As shown in Fig.~\ref{fig:model}(c,d), a higher frequency mode also results from Bragg diffraction of acoustic waves due to the patterning of holes along the beam's length. In the structure studied here the nominal hole parameters are chosen to be $a=2.23$~$\mu$m and $\Wx=\Wy=1.52$~$\mu$m, which results in a $100$~MHz phononic bandgap around a center frequency of $450$~MHz.  A defect is formed in the phononic lattice by increasing the hole width ($\Wx$) over the central $12$ holes of the beam, resulting in a localized ``breathing'' mode of frequency $\omega_\mathrm{m}/2\pi=458$~MHz that is trapped on either end by the phononic bandgap.  From the simulated motional mass of both mechanical resonances, the zero-point amplitude is estimated to be $\xzpf=8.1$~fm and $4.2$~fm for the flexural and breathing modes, respectively.

\begin{figure*}[btp]
\begin{center}
\includegraphics[width=1.0\textwidth]{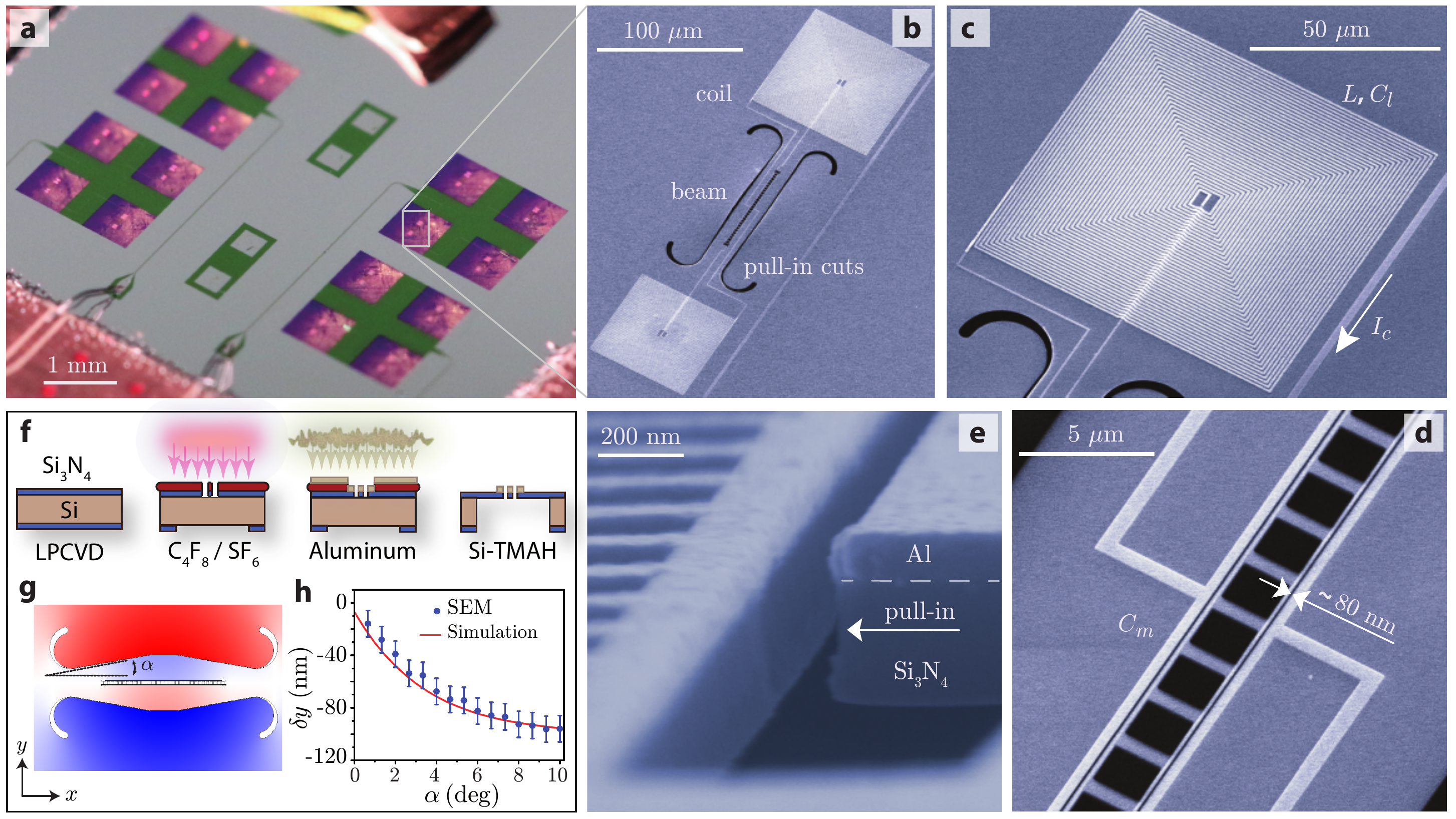}
\caption{\textbf{Sample fabrication.}
\textbf{a}, Optical image of the membrane microchip which is mounted, bonded and clamped to a low loss printed circuit board. The microchip contains four sets of four membranes.  In this image the $\SiN$ membranes of thickness $300~\mathrm{nm}$ are semi-transparent purple, the aluminum coated regions are gray, and the uncoated silicon substrate is green.  The two bright regions in the middle of each membrane correspond to the two coil resonators coupled to each nanobeam resonator. 
\textbf{b}, False color SEM image of the center part of the membrane depicting two aluminum planar coils (white) coupled to two sides of a single patterned phononic crystal nanobeam with stress pull-in cuts (black).
\textbf{c}, SEM image zoom-in of the spiral inductor ($p=1$~$\mu$m, $n=42$), showing the cross-overs needed to connect the inductor coil to the vacuum gap capacitor across the nanobeam resonator.
\textbf{d}, SEM image zoom-in of the released center region of the nanobeam mechanical resonator and vacuum gap capacitors with gap size of $s \approx 80$~nm.
\textbf{e}, Tilted SEM image of the capacitor gap showing the etch profile of the nanobeam and the aluminum electrode thickness ($\approx 65$~nm). 
\textbf{f}, Schematic of the main circuit fabrication steps: (i) LPCVD of stoichiometric $\SiN$ on both sides of a $200~\mu$m thick silicon substrate, (ii) $\mathrm{C}_4\mathrm{F}_8$:$\mathrm{SF}_6$ plasma etch through the nitride membrane defining the mechanical beam resonator and pull-in cuts on the top side, and membrane windows on the bottom side, (iii) electron beam lithography, aluminum deposition, and lift-off steps to pattern the microwave circuit, and (iv) final release of the nitride membrane using a silicon-enriched tetramethylammonium hydroxide (TMAH) solution. 
\textbf{g}, Simulation of the membrane relaxation during release.  The image shows the regions of positive (red) and negative (blue) displacement, $\delta y$, of the membrane. The stress release cuts (white) are shaped at an angle $\alpha$ to controllably narrow the capacitor gaps $s$ during release. The rounded shape of the pull-in cut end section has been optimized to minimize the maximal stress points to avoid membrane fracturing.
\textbf{h}, Plot of the simulated (solid red curve) and SEM-measured (blue solid circles) change in the slot gap ($\delta y$) versus slot-cut angle $\alpha$.  Error bars indicate the resolution limit in SEM measurements of the gap.} 
\label{fig:fab}
\end{center}
\end{figure*}

As motional capacitance scales roughly with mechanical resonator size, realizing large electromechanical coupling to nanomechanical resonators depends crucially on minimizing parasitic capacitance of the microwave circuit as per Eq.~(\ref{eq:coupling}).  Utilizing a planar spiral inductor coil of multiple turns greatly increases the coil inductance per unit length through mutual inductance between coil turns, and consequently, reduces coil capacitance.  One can determine the capacitance ($\Cl$) and inductance ($L$) of a given coil geometry by numerically simulating its self resonance frequency with and without a known small shunting capacitance.  Figure~\ref{fig:model}(f) displays a method of moments numerical simulation~\cite{SONNET} of the self resonance frequency ($\omegasrf$) of a series of square planar coil designs with constant area ($\Acoil = 87$~$\mu$m $\times 87$~$\mu$m) but varying wire-to-wire pitch $p$, or equivalently, coil turns $n$.  Here we assume a coil wire width and thickness of $500$~nm and $120$~nm, respectively, deposited on top of the $300$~nm nitride membrane.  While the coil capacitance is roughly constant at $\Cl \approx 2$~fF, the coil inductance varies over $3$ orders of magnitude, in good agreement with an analytical model for planar inductors~\cite{Mohan1999}.  An additional stray capacitance of $\Cs \approx 2.4$~fF is estimated for the full integrated microwave circuit (see App.~\ref{app:A} for details).   

%A finite element simulation of the pre-stressed nanobeam with $\sigma \sim 1$~GPa returns an $\omegam$ (from measurements). 

Figure~\ref{fig:model}(g) displays the simulated motional capacitance and vacuum coupling rate versus capacitor slot size $s$ for both the flexural and breathing modes of the beam resonator assuming a coil of pitch $p=1$~$\mu$m ($n=42$, $L=68$~nH, $\Cl=2.1$~fF, $\omegasrf/2\pi = 13.68$~GHz).  Here, $\partial C_\mathrm{m}/\partial u$ is calculated for each specific mechanical mode utilizing a perturbation theory depending on the integral of the electric field strength at the dielectric and metallic boundaries of the vacuum gap capacitor~\cite{Pitanti2015}.  For a gap size of $s=80$~nm, the vacuum coupling rate is estimated to be $\gzeroE/2\pi=58$~Hz ($240$~Hz for $\eta=1$) for the flexural mode and $\gzeroE/2\pi = 8$~Hz ($65$~Hz for $\eta=1$) for the breathing mode.  Note that here we assume the outer electrode of the vacuum gap capacitor extends along the entire length of beam in the case of the flexural mode, whereas for the breathing mode we limit the outer capacitor electrode to the central $6$ lattice constants of the beam where the breathing mode has significant amplitude.  Also, for the breathing mode simulations the two vacuum gap capacitors are assumed to be connected in parallel, which doubles the vacuum coupling rate due to the mode symmetry.     

Fabrication of the membrane microwave circuits begins with the LPCVD growth of $300$~nm thick stoichiometric $\SiN$ layers on the top and bottom surfaces of $200$~$\mu$m thick silicon wafer, and involves a series of electron beam lithography, dry etching, aluminum evaporation, and chemical wet etching steps.  An optical image of the fully fabricated and wirebonded chip is shown in Fig.~\ref{fig:fab}(a).  Zoom-in scanning electron microscope (SEM) images of the inductor coil and nanobeam regions of the device are shown in Fig.~\ref{fig:fab}(b-e). The main fabrication steps are depicted in Fig.~\ref{fig:fab}(f) and discussed in more detail in App.~\ref{app:C}. One important feature of our fabrication method is the use of the tensile membrane stress ($\sigma \approx 1$~GPa) to fabricate capacitive slot gaps that shrink upon release of the membrane, providing a controllable way to create ultra-small gaps.  As can be seen in the device figures of Fig.~\ref{fig:fab}(b,c), stress release cuts are used above and below the nanobeam region so as to allow the membrane to relax on either side of the beam.  A simulated plot of the membrane relaxation is shown in Fig.~\ref{fig:fab}(g).  Comparison of the simulated slot gap change ($\delta y$) and measured slot gap change for a series of fabricated devices with different cut angles $\alpha$ is shown in Fig.~\ref{fig:fab}(h), indicating that slot gap adjustments up to $100$~nm can be reliably predicted and produced.  In the measured device of this work we use this feature to controllably close the capacitor slot $s$ from an initial slot size of $s \approx 150$~nm right after dry etching, down to a final slot size of $s \approx 80$~nm after membrane release.

As shown in Fig.~\ref{fig:fab}(b), in the device studied here each nanobeam is coupled on one side to one coil, and on the other side to another coil.  The capacitor electrodes also extend across the whole length of the device in order to maximize coupling to the low frequency flexural mode of the beam.  The two coils have different lengths, resulting in different $LC$ resonant frequencies.  As will be presented elsewhere~\cite{Fink2015b}, such a double-coil geometry can be used to perform coherent microwave frequency translation using the intermediate nanomechanical resonator as a parametric converter~\cite{Hill2012,Lecocq2015}. In the following, however, we will focus on the lower frequency circuit (larger coil) only.  The device is cooled to a fridge temperature of $\Tf \approx 11$~mK using a cryogen free dilution refrigerator, and connected to a microwave test set-up consisting of low noise control and readout electronics for electromechanical characterization (see App.~\ref{app:B} for details). 

\section*{Coherent electromechanical response}
\label{sec:EIT}

\begin{figure*}[btp]
\begin{center}
\includegraphics[width=0.9\textwidth]{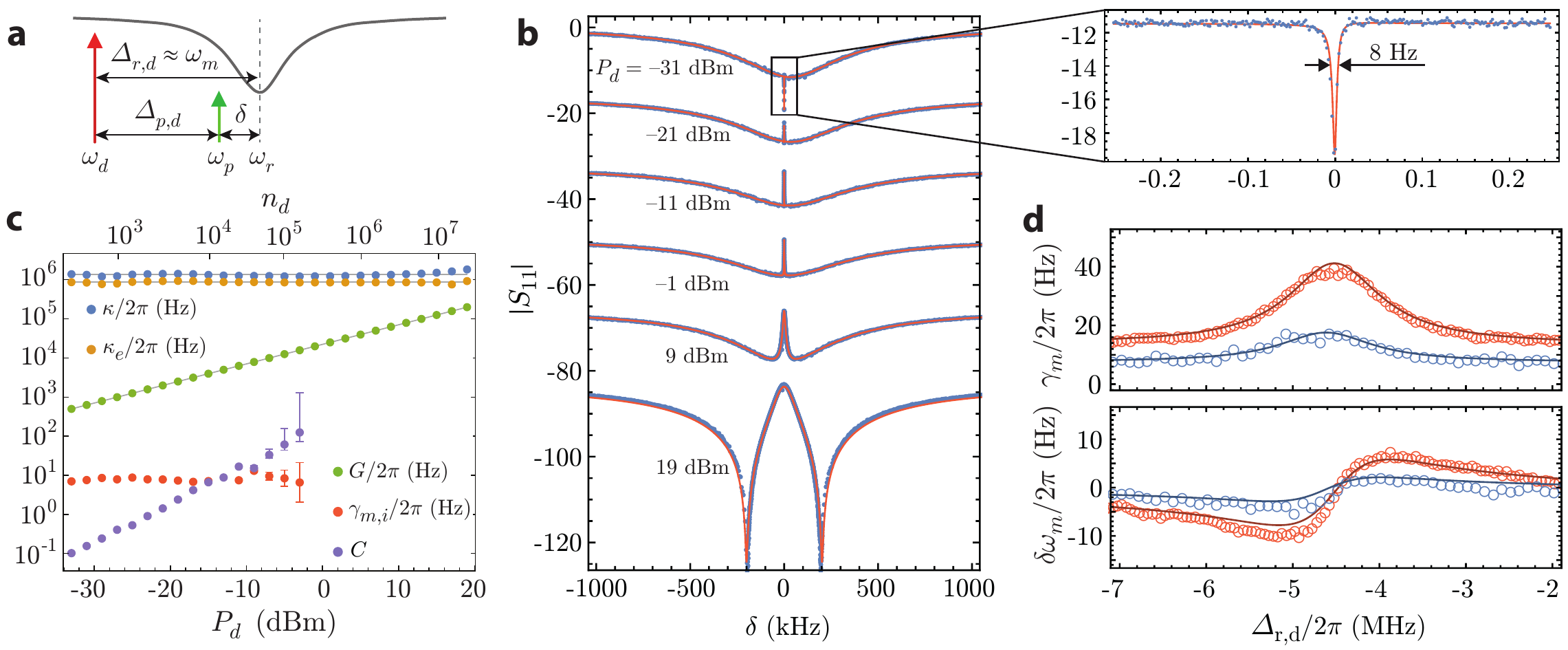}
\caption{\textbf{Coherent response. a,} Schematic of the two-tone EIT spectroscopy measurement.
\textbf{b}, Measured (blue points) probe spectra for different drive powers, all with a fixed drive detuning of $\Deltard \approx \omegam = 4.815$~MHz.  Each spectrum is offset by $-16.5$~dB for better visibility. Fits to measured spectra using Eq.~(\ref{eq:EIT}) are shown as solid red curves. Inset shows a zoomed-in view of the lowest power measurement with a mechanical linewidth of $\gammam/2\pi = 8$~Hz. 
\textbf{c}, Extracted system parameters (symbols) as a function of drive power using Eq.~(\ref{eq:EIT}) to fit the measured spectra.  Error bars correspond to a $95$~\% confidence interval in the fit to the measured spectra.
\textbf{d}, Mechanical linewidth $\gammam$ (top) and the mechanical frequency shift $\delta\omegam$ due to the optical spring effect (bottom) versus drive detuning $\Deltard$ at a fixed intra-cavity drive photon number.  Shown are the fit values from the measured probe spectra for two different fridge temperatures, $\Tf=11$~mK (blue circles) and $\Tf=114$~mK (red circles).  The drive photon number at $\Tf=11$~mK ($\Tf=114$~mK) is equal to $\ncavd = 2350$ ($5980$).  The solid line curves are a fit to the damping and spring shift using a radiation pressure back-action model as per Ref.~\cite{Teufel2008}.}
 \label{fig:EIT}
\end{center}
\end{figure*}

Sweeping a narrowband microwave source across the $6$~GHz-$12$~GHz frequency range, and measuring in reflection, we find a high-$Q$, strongly coupled microwave resonance at $\omegacE/2\pi=7.965$~GHz corresponding to the larger coil of $42$ turns.  This is very close to the expected $LC$ resonance frequency based upon the above simulations, indicating that the stray and motional capacitance of the circuit are close to the expected values.  Using a two-tone pump and probe scheme we are able to study the coherent interaction between the microwave electrical circuit and the coupled nanobeam mechanical resonator.  In the driven linearized limit~\cite{Aspelmeyer2014}, the circuit electromechanical system is approximately described by an interaction Hamiltonian $\HOM = \hbar \GOM (\adag \bhat + \ahat \bdag)$, where $\ahat$ ($\adag$) is the microwave photon annihilation (creation) operator for the $LC$ resonator mode of the circuit and $\bhat$ ($\bdag$) are the phonon annihilation (creation) operators of the mechanical resonance. $\GOM = \gzeroE \sqrt{\ncavd}$ is the parametrically enhanced electromechanical coupling strength, with $\ncavd$ corresponding to the number of intra-cavity microwave drive photons inside the resonator.  As schematically indicated in Fig.~\ref{fig:EIT}, pumping with a strong drive at a detuning $\Deltard \equiv \omegacE - \omegad \approx \omegam$ from the $LC$ resonance of the circuit produces a two-photon resonance condition with a second (weaker) probe tone as it is swept across the microwave resonance.  Interference in the reflected probe signal occurs between that part of the probe field which enters the microwave resonator and is directly re-emitted, and that part of the probe field which enters the cavity, interacts with the mechanical resonator, and is then re-emitted from the cavity. Probing this interference as a function of the probe detuning $\delta$ yields the optomechanical analog of electromagnetically induced transparency (EIT)~\cite{Weis2010,Safavi-Naeini2011,Teufel2011}.  

For red-sideband pumping ($\Deltard \approx \omegam$) the expected probe reflection spectrum is given by, 

\begin{equation}
S_{11}(\delta) = 1- \frac{\kappaEe}{\kappaE/2 + i\delta + \frac{2\GOM^2}{\gammai + i2(\delta-(\omegam-\Deltard))}},    
\label{eq:EIT}
\end{equation}

\noindent where $\delta \equiv \omegap - \omegacE$ is the detuning of the probe frequency ($\omegap$) from the cavity resonance ($\omegacE$), $\kappaEe$ is the external microwave cavity damping rate due to coupling to the coplanar waveguide port, $\kappaEi$ the intrinsic cavity damping rate, and $\kappaE=\kappaEi + \kappaEe$ is the total loaded cavity damping rate.  Here we have made approximations assuming the system is sideband resolved ($\omegam/\kappaE \gg 1$) and that the probe signal is weak enough so as to not saturate the drive tone.  The cooperativity associated with the coupling of the microwave cavity field to the mechanical resonator is given by $C \equiv 4\GOM^2/\kappaE\gammai$, where $\gammaOME = 4\GOM^2/\kappaE$ is the back-action-induced damping of the mechanical resonator by the microwave drive field.

In this work we focus on the fundamental in-plane flexural mode of the beam.  The phononic crystal breathing mode at a frequency of $450$~MHz is not accessible in our current single microwave resonator circuit given the high drive power required to excite the circuit at the large cavity detuning required for two-photon resonance.  In future work a double resonant system~\cite{Dobrindt2010} may be employed to overcome this limitation and allow for efficient excitation and detection of high frequency mechanical resonators such as the breathing mode.  By stepping the pump detuning frequency ($\Deltard$) and sweeping the probe signal across the cavity resonance, an EIT-like transparency window in the microwave cavity response is found at a drive detuning of $4.4815$~MHz, close to the theoretically simulated resonance frequency ($4.18$~MHz) of the fundamental in-plane flexural mode.  Figure~\ref{fig:EIT}(b) shows a series of measured probe spectra (blue points) at different applied drive powers for a drive detuning fixed close to the two-photon resonant condition of $\Deltard/2\pi \approx \omegam = 4.48$~MHz.  Fits to the measured spectra are performed using Eq.~(\ref{eq:EIT}) and plotted as solid red curves in Fig.~\ref{fig:EIT}(b). 

From each fit we extract the loaded microwave resonator properties ($\kappaE$, $\kappaEe$, $\omegacE$), the parametric coupling rate ($\GOM$), the mechanical frequency ($\omegam$), and the intrinsic mechanical damping rate ($\gammai$).  These fit values are plotted versus drive power in Fig.~\ref{fig:EIT}(c).  The microwave cavity parameters ($\kappaE/2\pi \approx 1.28$~MHz, $\kappaEe/2\pi \approx 0.896$~MHz) are found to be approximately constant over $5$ orders of magnitude in drive power, up to an intra-cavity photon number of $\ncavd \approx 2\times10^{6}$.  For $\ncavd \gtrsim 2\times 10^{6}$ the intrinsic damping of the cavity begins to rise, and above $\ncavd \approx 4\times 10^{7}$ ($\Pd = 19$~dBm) the $LC$ circuit goes normal.  Conversion from drive power to intra-cavity photon number $\ncavd$ is performed using the thermometry calibrations described in the next section.  At low drive power ($C \lesssim 100$) the fits yield high confidence estimates of both $C$ and $\gammai$, with the intrinsic mechanical damping of the resonator estimated to be $\gammai/2\pi = 8$~Hz at the lowest drive powers (see inset to Fig.~\ref{fig:EIT}(b)).  At high drive powers ($C \gtrsim 100$) the transparency window saturates and becomes too broad to accurately determine either $C$ or $\gammai$.  As such we only provide fit estimates for $C$ and $\gammai$ below a cooperativity of $100$.  

%For all drive powers, however, $\GOM$ can be accurately estimated from the fits.  From the calibrated intra-cavity photon number and a linear fit to the measured back-action rate versus drive strength we estimate the vaccum electromechanical coupling rate to the $4.4815$~MHz flexural mode to be $\gzeroE/2\pi = 41.5 \pm 0.6$~Hz.

%Also, spectral diffusion of the mechanical resonance can lead to a broadened and reduced amplitude of the transparency dip (or peak) at low drive power, leading to an overestimate of $\gammai$ and an underestimate of $C$.  

Figure~\ref{fig:EIT}(d) shows a plot of the measured mechanical frequency shift ($\delta\omegam$) and damping ($\gammam \equiv \gammai + \gammaOME$) versus drive detuning $\Deltard$.  Here we adjust the drive power as a function of drive detuning so as to maintain a constant intra-cavity drive photon number, and fit the transparency window using a Fano lineshape (see SI).  Data was taken at $\Tf=11$~mK as well as at an elevated fridge temperature of $\Tf=114$~mK.  The intra-cavity drive photon number in both cases was chosen to yield a peak cooperativity of order unity.   We observe broadening of the mechanical linewidth that peaks at a detuning $\Deltard$ equal to the mechanical resonance frequency, and stiffening (softening) of the mechanical mode for drive detuning above (below) the mechanical resonance frequency.  Plots of the theoretical damping and frequency shift due to radiation pressure backaction~\cite{Teufel2008} are shown as solid back curves in Fig.~\ref{fig:EIT}(d).  We find a parametric coupling rate $G/2\pi = 1.80$~kHz ($2.98$~kHz) and intrinsic mechanical damping rate $\gammai/2\pi = 7.7$~Hz ($14$~Hz) that fit both the damping and spring shift curves at $\Tf=11$~mK ($114$~mK), in close agreement with the estimated values from the fixed detuning data in Fig.~\ref{fig:EIT}(b). 

\section*{Mode thermometry and backaction cooling}
\label{sec:GSC}

\begin{figure*}[btp]
\begin{center}
\includegraphics[width=0.9\textwidth]{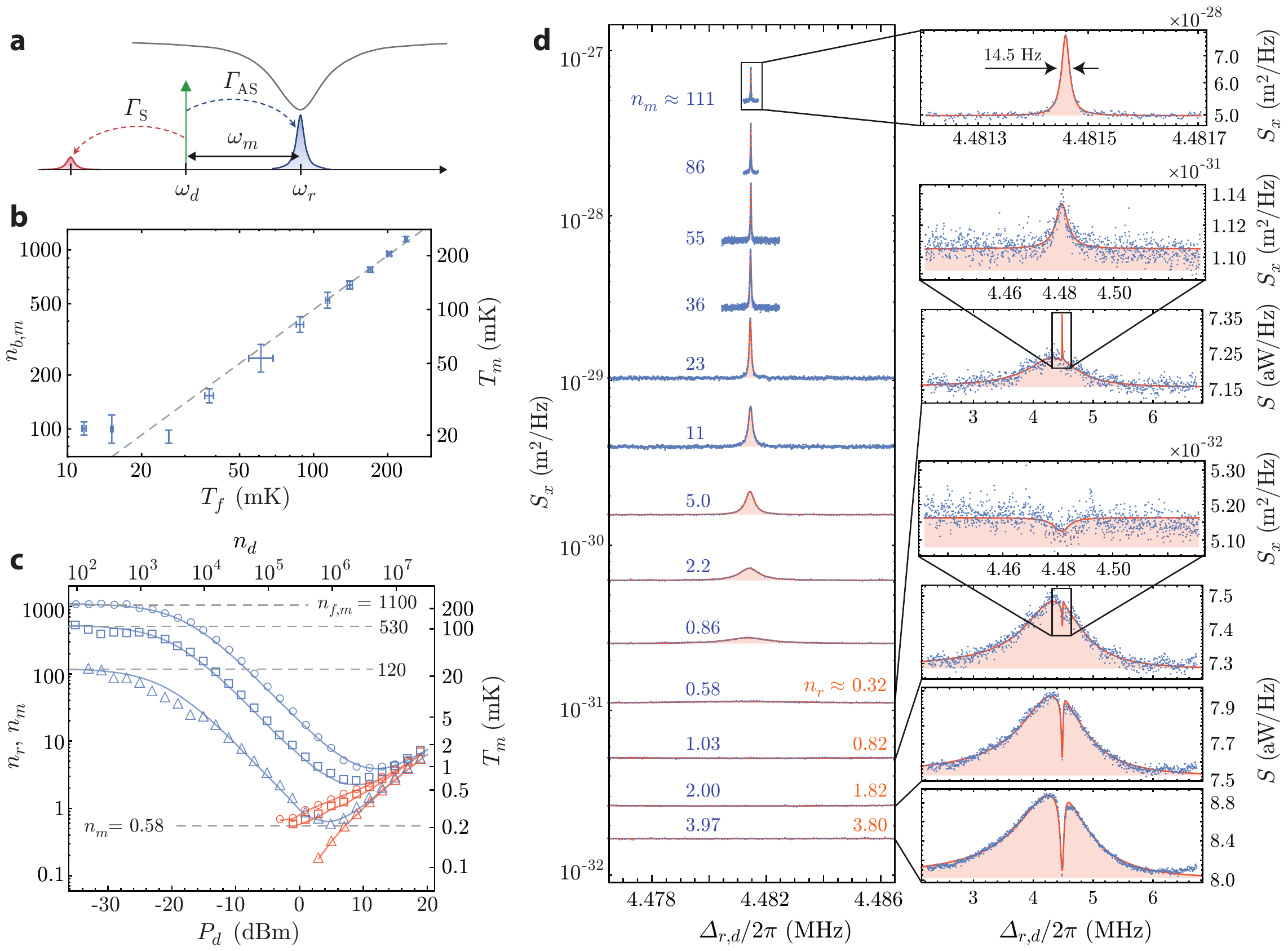}
\caption{\textbf{Mechanical displacement noise a,} Schematic showing the pump detuning and scattered microwave signals used to measure the mechanical resonator's displacement noise.
\textbf{b}, Plot of the measured mechanical resonator bath phonon occupation ($\nmb$) and effective temperature ($\Tm$) as a function of the fridge temperature ($\Tf$).  Each data point corresponds to the average inferred occupancy for blue detuned driving ($\Deltard = -\omegam$) at low cooperativity ($C \ll 1$).  Error bars corresponding to the standard deviation in the inferred occupancy over several temperature sweeps.  Calibration in units of occupancy is performed using a fit to Eq.~(\ref{eq:calibration}) as described in the main text and SI. 
The gray dashed lines show the expected Bose-Einstein distribution ($\nbar=( \exp{(\frac{\hbar \omegam}{k_\mathrm{B} \Tf})-1} )^{-1}$) assuming perfect thermalization to the fridge.
\textbf{c}, Plot of the dynamic backaction cooling of the mechanical resonator versus drive power at three different fridge temperature: $\Tf=235$~mK ($\nfm = 1100$; open circles), $\Tf=114$~mK ($\nfm = 530$; open squares), and $\Tf=26$~mK ($\nfm = 120$; open triangles). Data points showing the estimated average phonon occupancy of the fundamental in-plane flexural mode at $\omegam/2\pi \approx 4.48$~MHz ($\nbar$) are shown as blue symbols, whereas data points for the estimated microwave cavity photon occupancy at $\omegacE/2\pi=7.498$~GHz ($\nr$) are shown as orange symbols.   The corresponding effective mode temperature, $\Tm$, of the flexural mechanical mode is also shown on the right vertical axis.  The solid line blue curves correspond to a model for the expected mechanical mode occupancy using the a fit to the measured drive power relation for $\GOM$ and the microwave cavity parameters from coherent two-tone spectroscopy, the intrinsic mechanical damping from low power thermometry measurements, and a fit to the power dependence of the microwave resonator occupancy.    
\textbf{d}, Measured anti-Stokes noise displacement spectrum for several different drive powers at $\Tf = 26$~mK (blue data points).  Fits to the measured spectra are shown as red solid lines (see App.~\ref{app:F} for fit model).  Extracted values for $\nbar$ and $\nr$ are indicated and correspond to the results presented in panel (c). Zoom-ins of the cavity noise and measured noise peaks are shown as insets.} 
\label{fig:thermal}
\end{center}
\end{figure*}

Measurement of the mechanical resonator noise is used to calibrate the delivered microwave power to the circuit and to study the backaction cooling of the mechanical resonator.  In the resolved sideband limit ($\omegam/\kappa \gg 1$), efficient scattering of drive photons by mechanical motion occurs for $\Deltard = \pm\omegam$, in which either anti-Stokes ($\Deltard = \omegam$) or Stokes ($\Deltard=-\omegam$) scattering is resonant with the cavity.  Blue detuned pumping at $\Deltard = -\omegam$ results in Stokes scattering of the drive field, down-converting a photon to the cavity resonance and emitting a phonon into the mechanical resonator in the process.  Red detuned pumping at $\Deltard = \omegam$, as illustrated in Fig.~\ref{fig:thermal}(a), leads to predominantly anti-Stokes scattering in which a drive photon is up-converted to the cavity resonance and a phonon is absorbed from the mechanical resonator.  The per-phonon anti-Stokes scattering rate for this pumping geometry is $\GammaAS \approx 4G^2/\kappa$, to a good approximation equal to the backaction damping rate $\gammaOME$ which leads to cooling of the mechanical resonator~\cite{Marquardt2007}.  

Figure~\ref{fig:thermal}(b) shows a plot of the measured area underneath the Lorentzian noise peak of the fundamental in-plane mechanical resonance versus fridge temperature.  Here, data for blue detuned ($\Deltard=-\omegam$) driving has been averaged over several different temperature sweeps, with the area at each temperature normalized to units of phonon occupancy ($\nmb$) using the high temperature measurement ($\Tf = 235$~mK) as a reference point.  In these measurements the drive power was kept at a low enough value to ensure $C \ll 1$ and negligible backaction damping or amplification.  The mechanical flexural mode is seen to thermalize with the fridge temperature all the way down to $\Tf \approx 25$~mK, at which point the mechanical mode temperature saturates.  The source of this temperature saturation in the mechanics is not fully understood, but is thought to be due to coupling to two-level systems (TLS) in the amorphous $\SiN$ membrane~\cite{Sarabi2015b}.  These TLS can be driven by the microwave input signal into an elevated temperature state, and, as presented in the SI, can also strongly couple with the high impedance microwave cavity resonance.  This latter property may interfere with the mechanical transduction process, leading to unreliable thermometry of the mechanical mode.  

For a known temperature of the mechanical resonator, one may also employ the above low-cooperativity thermometry measurement to calibrate the vacuum coupling rate $\gzeroE$ between the mechanics and the microwave circuit~\cite{TeufelPrivate} (see App.~\ref{app:H} for details).  As the reflected drive signal and the scattered photons by the mechanical mode experience the same amount of gain, normalizing the measured reflected noise spectrum ($S(\omega)$) by the measured reflected drive tone amplitude ($\Pref$) yields a Lorentzian of the following form for a drive detuning of $\Deltard=\omegam$,

\begin{equation}
\frac{S(\omega)}{\Pref} \approx  \mathcal{O} + \frac{16 \gzeroE^2 \kappaEe^2}{\left((\kappaE - 2\kappaEe)^2+4\omegam^2\right)\left(\kappaE^2+4(\omegam-\omega)^2 \right)} \times \frac{4 \nmb \gammai}{\gammai^2+4(\omega-\omegam)^2}.
\label{eq:calibration}
\end{equation}

\noindent The background offset $\mathcal{O}$ yields the added noise of the measurement amplifier chain; $\nadd \approx 30$ for our current set-up.  Integrating the normalized spectral density for the reference fridge temperature of $\Tf=235$~mK ($\nfm = 1100$), and assuming $\nmb = \nfm$, yields a vacuum coupling rate of $\gzeroE/(2\pi)=41.5$~Hz, comparable to that estimated from numerical simulation ($58$~Hz for $s=80$~nm).  With $\gzeroE$ calibrated, the conversion factor between drive power and intra-cavity drive photon number can now be determined from the coherent two-tone spectroscopy measurements of $\GOM = \gzeroE\sqrt{\ncavd}$, as displayed in Fig.~\ref{fig:EIT}(c).     

Increasing the drive power to large cooperativity levels results in backaction cooling of the mechanical resonator for detuning $\Deltard=\omegam$.  Figure~\ref{fig:thermal}(c) plots the measured occupancy of the mechanical resonator versus drive power for three different fridge temperatures, $\Tf=235$, $114$, and $26$~mK.  For the lowest of these temperatures ($\Tf=26$~mK), the measured noise power spectral density from low to high drive power are shown in Fig.~\ref{fig:thermal}(d).  At low drive powers we find excellent agreement between the inferred $\nbar$ and the bath occupancy corresponding to the fridge temperature, $\nfm$, for all three temperatures. At intermediate drive powers the mechanical mode is both damped and cooled according to $\nbar=\nfm/(C+1)$.  At the highest drive powers we measure both an increase in broadband added noise and Lorentzian microwave cavity noise.  We attribute this excess noise to absorption of the input microwave drive.  These two additional noise inputs can lead to noise squashing in the measured output spectrum and heating of the mechanical resonator~\cite{Dobrindt2008,Rocheleau2010,Teufel2011,Safavi-Naeini2013a}. Using a model that includes microwave (thermal) noise in the input line ($\nwb$) and in the microwave cavity ($\nrb$) (see App.~\ref{app:F} for details), we fit the measured spectra at higher drive power for the mechanical mode occupancy $\nbar$ (blue symbols) and the microwave cavity noise occupancy $\nr$ (orange symbols).  The lowest mechanical occupancy is found to be $\nbar = 0.58$ for a drive photon number of $\ncavd = 10^6$ and a fridge temperature of $\Tf=26$~mK, and is similar to the lowest occupancies realized to date for other backaction cooled electromechanical resonators~\cite{Teufel2011,Wollman2015,Pirkkalainen2015}.  Measurements at the lowest fridge temperature of $\Tf = 11$~mK resulted in inconsistent and fluctuating cooling curves, attributable we believe to drive-power-dependent coupling of individual TLS to the microwave cavity (see App.~\ref{app:I} for details). 

\section*{Discussion and Outlook}
\label{sec:conc}

Utilizing $\SiN$ nanomembranes as a substrate for superconducting microwave circuits enables the formation of high-impedance circuit elements with large per photon electric field strengths.  In the current work, the reduced thickness and low dielectric constant of the nanomembrane helps realize a microwave resonator with an estimated vacuum field strength as large as $E_{\text{vac}} \approx 260$~V/m.  This feature gives rise to the large electromechanical coupling that we observe to the fundamental flexural mode of an integrated phononic crystal nanobeam.  Dynamical backaction cooling via a strong microwave drive tone results in an occupancy of $\nbar = 0.58$ for the $4.48$~MHz flexural mode of the beam, limited here by heating of the circuit due to absorption of the microwave drive at the highest powers.  Substantial further reduction in the coil and stray capacitance should be possible through tighter coil wiring and optimized layout of the capacitor wiring, respectively, greatly reducing the required drive power (and corresponding heating) for backaction cooling to the quantum ground state.  

Our results also indicate that capacitive coupling to smaller, much higher frequency nanomechanical resonant modes is possible utilizing the $\SiN$ platform.  In particular, the planar nature of the membrane circuit allows for integration with slab phononic crystals which can be used to guide and localize mechanical excitations over a broad ($100$~MHz-$10$~GHz) frequency range~\cite{Safavi-Naeini2010b}.  Numerical simulations show that the electromechanical coupling strength to the localized ``breathing'' mode at $458$~MHz of the phononic crystal nanobeam of our device is large enough for efficient photon-phonon parametric coupling; however, a doubly-resonant microwave cavity is needed to drive the system in such a deeply sideband resolved limit~\cite{Dobrindt2010}.  Coupling phononic crystal structures to superconducting microwave circuits would allow not only for exquisite studies of phonon dynamics using the toolbox of circuit-QED~\cite{Gustafsson2014}, but could also be used to realize a quantum network involving superconducting qubits, phonons, and optical photons~\cite{Regal2011,Safavi2011,Barzanjeh2011,Habraken2012}.  

\begin{acknowledgments}The authors would like to thank Joe Redford, Lev Krayzman, Matt Shaw, and Matt Matheny for help in the early parts of this work. LH thanks Andreas Wallraff for his support during his Master's thesis stay at Caltech.  This work was supported by the DARPA MESO program, the Institute for Quantum Information and Matter, an NSF Physics Frontiers Center with support of the Gordon and Betty Moore Foundation, and the Kavli Nanoscience Institute at Caltech. AP was supported by a Marie Curie International Outgoing Fellowship within the $7^{th}$ European Community Framework Programme, NEMO (GA 298861).  
\end{acknowledgments} 

%\bibliography{../JMFandOJPbib}

\appendix

\section{Circuit properties}
\label{app:A}  
\subsection{Coil simulation}
Our device is fabricated and simulated on a 300~\textrm{nm} thick and $(777~\mu\mathrm{m})^2$ large $\mathrm{Si}_3\mathrm{N}_4$ membrane. The coil wire is 500~\textrm{nm} wide and 120~\textrm{nm} thick, with a $1~\mu$m pitch, 42 turns forming a square with lateral length of only $87~\mu\mathrm{m}$, 
well in the lumped element limit. According to finite element simulations, which includes wire cross-overs, the coil is inductive up to its self resonance frequency of $\nu_\mathrm{srf}=13.38$~GHz, where the half wavelength roughly matches the total wire length of $l=7.7$~mm. We repeat this simulation with a small additional shunt capacitor of known value ($\Delta C=0.1$~fF) and extract the new self resonance frequency $\nu_\mathrm{srf,2}$. Solving the two simple relations $\omega_\mathrm{srf}=(L C_\mathrm{l})^{-1/2}$ and $\omega_\mathrm{srf,2}=(L (C_\mathrm{l}+\Delta C))^{-1/2}$, we extract $L=68~\textrm{nH}$ and $C_\mathrm{l}=2.1~\mathrm{fF}$. These results are valid close to - but below - the self resonance frequency of the coil. In this limit we realize a maximum impedance of $Z_0=\sqrt{L/C_\mathrm{l}}\approx 5.7~\mathrm{k}\Omega$, far exceeding the vacuum impedance $Z_\mathrm{vac}\approx 377~\Omega$, and approaching the resistance quantum $R_\mathrm{q}=h/(2e)^2\approx 6~\mathrm{k}\Omega$. 

\subsection{Full circuit parameters}
Knowing the inductance $L$ of the fabricated inductor, as well as the actually measured resonance frequency of $\omega_\mathrm{r}/(2\pi)=7.965~\mathrm{GHz}$, yields a total capacitance of $C_\mathrm{tot}=C_\mathrm{l}+C_\mathrm{m}+C_\mathrm{s}=5.87~\mathrm{fF}$ and a total circuit impedance of $Z_\mathrm{tot}=3.4~\mathrm{k}\Omega$ (see Fig.~\ref{fig:setup}). The modulated capacitance $C_\mathrm{m}$ is a function of the capacitor slot size, which we estimate from SEM images to be on the order of $s\approx 80-100~\mathrm{nm}$. Numerical finite element simulations yield a nanobeam capacitance of $C_\mathrm{m}\approx 1.4~\mathrm{fF}$ for this gap size (see Figs. in main text), which gives a participation factor of $\eta\approx 0.25$. 

Using a self resonance frequency simulation of the full electrical circuit including $C_\mathrm{m}$, we can attribute the remaining stray capacitance of $C_\mathrm{s}\approx 2.4~\mathrm{fF}$ to the coil to capacitor wiring ($57$~\%), the presence of a second resonant circuit ($14$~\%), the coupling wire ($7$~\%), non-ideal crossovers ($7$~\%). The remaining 0.35~\textrm{fF} ($15$~\%) we attribute to frequency dependence, packaging and our uncertainty of the relative permittivity of silicon nitride at low temperature $\epsilon_\mathrm{r}\approx 8$. As expected, for these full circuit simulations we extract the same inductance $L$ as for the coil only simulations. The value of $L=68~\textrm{nH}$ is consistent with both, the modified Wheeler and the current sheet method \cite{Mohan1999}, to within $\pm 2$~nH.

\subsection{High frequency mechanical mode}
\begin{figure}[h]
\begin{center}
\includegraphics[width=0.7 \columnwidth]{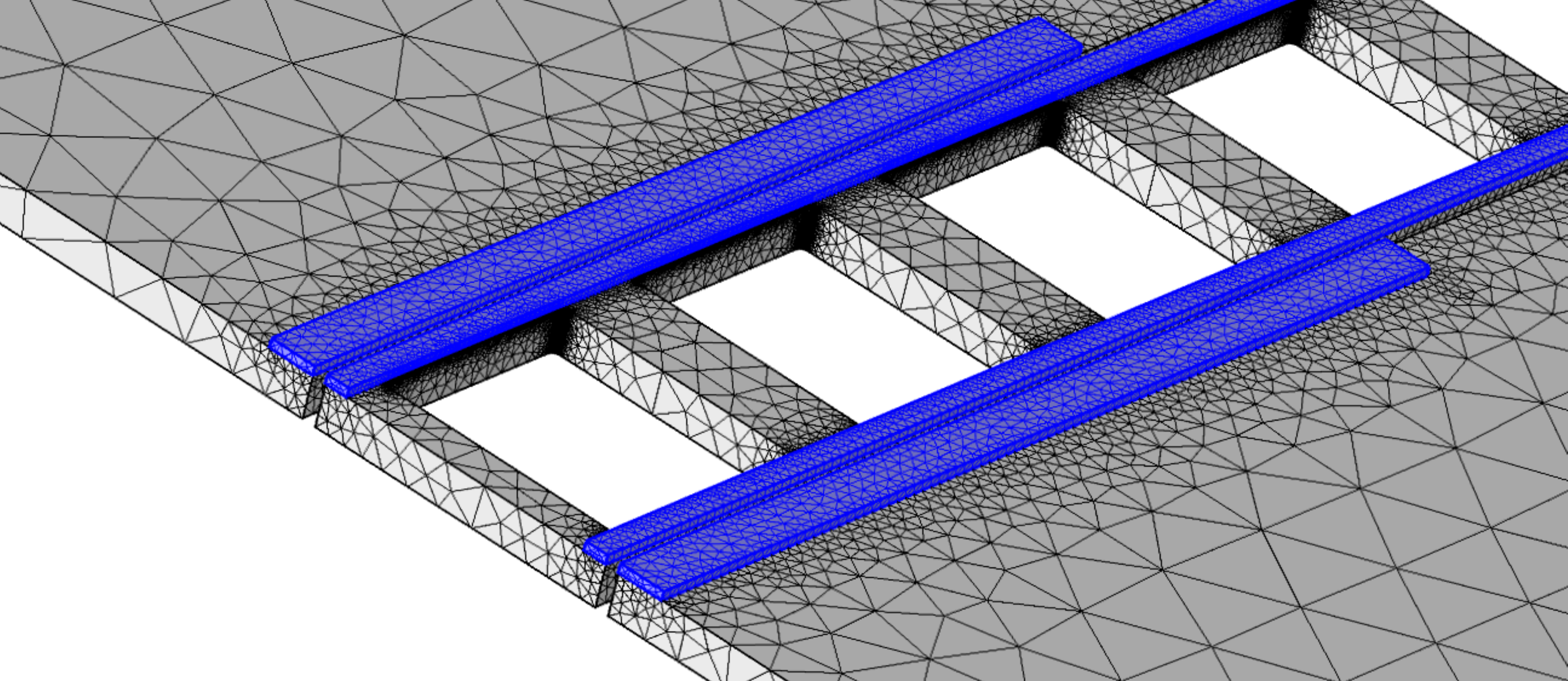}
\caption{\textbf{Geometry for high frequency breathing mode simulation.}
The simulated structure with 65 nm thick Aluminum (highlighted blue) on 300 nm thick silicon nitride (gray) is shown. We use a symmetric boundary condition in the center of the beam (bottom left).} 
\label{fig:simulation}
\end{center}
\end{figure}

In order to estimate the electromechanical coupling of the high frequency acoustic mode, we consider that the identical microwave circuit is coupled to both sides of the nanobeam. Here the outer capacitor length is taken to match the acoustic defect region of $2\times3$ lattice constants, see Fig.~\ref{fig:simulation}. We find a reduced participation ratio $\eta = 0.11$ due to the reduced $C_\mathrm{m}$ in this case. Further improvements in reducing the circuit's stray capacitance will have a big impact for efficient coupling to high frequency modes.

\section{Experimental setup}
\label{app:B}
\begin{figure}[h]
\begin{center}
\includegraphics[width=0.65 \columnwidth]{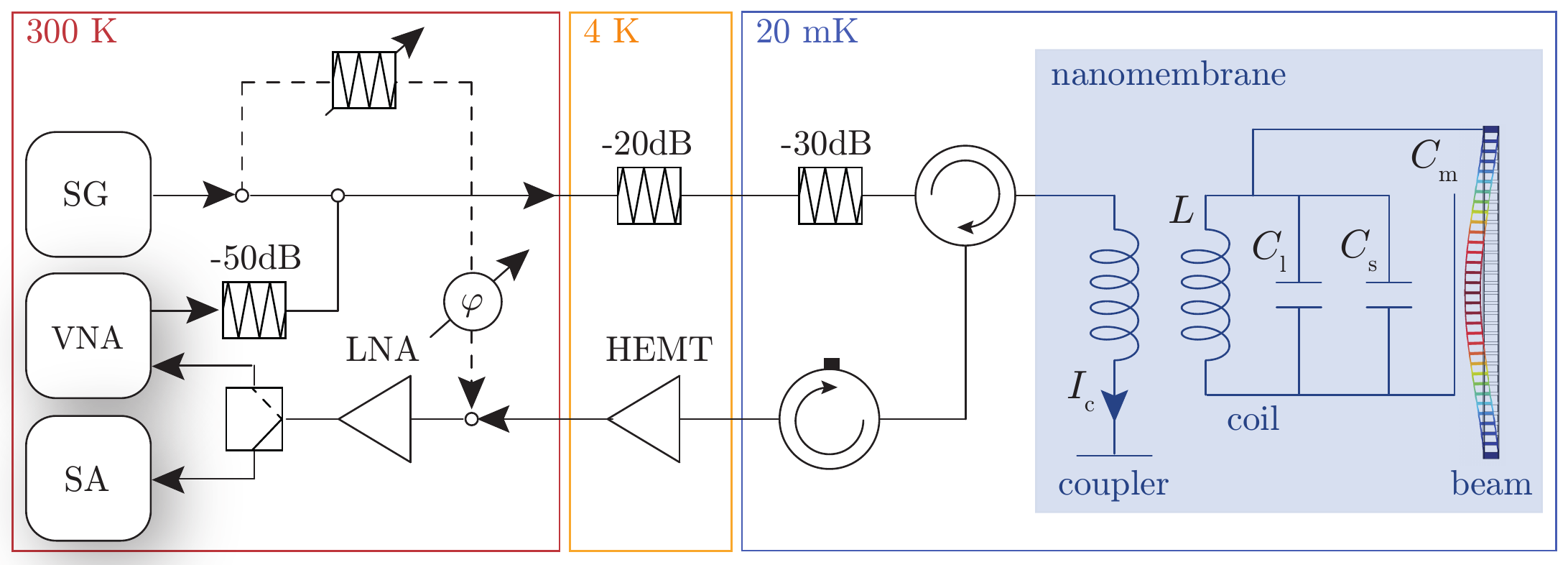}
\caption{\textbf{Experimental setup.}
The output tone of a microwave signal generator (SG) and the output tone of a vector network analyzer (VNA) are combined at room temperature, attenuated, routed to the sample at about 20~mK and inductively coupled to the LC circuit on the nanomembrane. We detect the reflected output tone after amplification with a high electron mobility transistor amplifier (HEMT), switchable pump tone cancelation (dashed lines), and further amplification with a low noise amplifier (LNA). The measurement is done either phase coherently with the VNA, or we detect the incoherent power spectrum with the spectrum analyzer (SA).} 
\label{fig:setup}
\end{center}
\end{figure}

%\subsection{Input}
For the measurements of coherent and incoherent electromechanical response, we combine the output of a vector network analyzer with up to two microwave sources, feed the microwave tones to the base plate of a cryogen free dilution refrigerator using UT-085 stainless steel coaxial cables with feedthroughs for thermalization at each temperature stage and an additional attenuation of 50~dB to suppress room temperature Johnson noise (see Fig.~\ref{fig:setup}). We couple to the sample in a reflective geometry using a circulator and a low loss, high dielectric constant, copper printed circuit board (PCB). On the PCB and chip we use 50~$\Omega$ coplanar waveguides to route the microwave tones all the way to the membrane with very little reflections. On the membrane the center conductor is shorted to ground through a small wire, which couples the waveguide inductively to the LC resonant circuit. 

%\subsection{Output}
On the output side we use another circulator for isolating the sample from 4~K noise, otherwise entering in reverse direction. Crimped niobium titanium superconducting cables are used to connect directly to a low noise, high electron mobility transistor amplifier (HEMT) at 4~K. From there we use loss loss UT-085 stainless steel - beryllium copper cables and amplify once more at room temperature. In order to suppress spurious response peaks in high drive power cooling measurements, we add a phase and amplitude adjusted part of the pump tone to the output signal, as shown in Fig.~\ref{fig:setup}. Properly adjusted, this cancels the directly reflected pump tone before entering the low noise amplifier (LNA) and spectrum analyzer (SA) and avoids the occurrence of spurious resonances in the measured spectrum. After the final amplification we use an electronically controlled microwave switch to distribute the signal to either the spectrum analyzer or the second vector network analyzer port.

\section{Device Fabrication}
\label{app:C}
\subsection{Wafer preparation}
After a thorough RCA clean we grow a 300~\textrm{nm} thick film of stoichiometric Si$_3$N$_4$ at a temperature of 835~\textrm{C}, using low pressure chemical vapor deposition on both sides of a doubly polished 200~$\mu\mathrm{m}$ thick, high resistivity ($>10$~k$\Omega$-cm), Si $\langle 100 \rangle$ wafer. After cooldown, the dielectric film has a stress of $\approx 1$~GPa due to the differential expansion coefficient. We spin a protective layer of photoresist and dice the wafer in $10$~mm $\times 10$~mm chips. 

\subsection{Membrane patterning}
The chips are cleaned using weak sonication in ACE and IPA and prebaked at 180~$^{\circ}$C for 2 min on a hotplate. We then spin the front side with ZEP 520A at 4000~rpm (for protection), bake at 180~$^{\circ}$C for 2 min, spin the back side with ZEP 520A at 2000~rpm and bake at 180~$^{\circ}$C for 2 min. Patterning of the 16 membrane areas of size $1$~mm $\times 1$~mm each, is done with 100~keV electron beam exposure with a 200~nA electron beam, 50~nm fracturing size and a dose of 250~$\mu\mathrm{C}/\mathrm{cm}^2$ on the chip back side. This layer is carefully aligned to the chip corners. We develop with ZND-N50  for 2.5~min and rinse in MIBK for 0.5~min. This is followed by an ICP-RIE etch of the silicon nitride in the developed areas, using a C$_4$F$_8$/SF$_6$ (34/12~sccm) plasma, generated with an ICP power of 1000~W, RF power of 30~W and a DC bias of 84~V, at a pressure of 15~\textrm{Torr} and a temperature of 25~$^{\circ}$C for 7~min 15~s. We finish this layer by a thorough cleaning of the chips using weak sonication in TCE, IPA, ZDMAC, ACE and IPA.

\subsection{Nanobeam patterning and membrane pre-etching}
This layer initially follows the same procedure to pattern the top side of the chip (no resist on the back side) with the nanobeams, pull-in cuts and the global and pattern alignment markers of size $(20\time20~\mu\mathrm{m})^2$, with these process parameters: 300~pA beam, 2.5~nm fracturing, 275~$\mu\mathrm{C}/\mathrm{cm}^2$ dose, 7~min 50~s etch time. We then use an o-ring sealed holder to expose only the back side of the chip to $30$~\% KOH in water at 85~$^{\circ}$C (stir bar at 400~rpm). This anisotropic Si wet etch is stopped when the wafer becomes semi-transparent (dark orange) in the membrane area, when illuminated with an LED on the sealed side of the chip. The color indicates a silicon thickness of $\approx 5~\mu\mathrm{m}$ which is usually achieved after $2.5~$ h of etching. After cleaning the chip in ultra-pure deionized water and IPA, we wet etch both the front and back side of the chip in $30$~\% KOH in water at 65~$^{\circ}$C (stir bar at 100~rpm) for $70~$s. This partially undercuts ($\approx 100~\mathrm{nm}$) the nanobeams for a clean subsequent inverse shadow evaporation process \cite{Pitanti2015}, used to pattern the small gapped capacitors. The chips are then rinsed in hot water, fresh piranha solution (mix 45~mL H$_2$SO$_4$ with 15~mL H$_2$O$_2$ at 85~$^{\circ}$C with stir bar at 300~rpm) for 8~min followed by a water and IPA rinse.

\subsection{Capacitors and ground plane}
This layer patterns all of the electrical circuit, except for the coil wires. We start with a prebake at 180~$^{\circ}$C for 2 min, and spin the front side with ZEP 520A at 2000~rpm, followed by another bake at 180~$^{\circ}$C for 2 min. We use 100~keV electron beam lithography to pattern the ground plane and transmission lines (200~nA beam, 50~nm fracturing, 290~$\mu\mathrm{C}/\mathrm{cm}^2$ dose with PEC), as well as the capacitor wires, and the wires connecting the capacitors with the coil end and center  (10~nA beam, 10~nm fracturing, 275~$\mu\mathrm{C}/\mathrm{cm}^2$ dose). This layer is carefully aligned to the etched negative markers from the previous step. We develop the chips in the same way and use a $O_2$ plasma ash process (50~sccm O$_2$, 0.74~bar, 13.56~MHz, 35~W, 2~min) to descum the surface before deposition of aluminum. For the deposition we use an electron beam evaporator (0.3~nm/s, 65 nm thickness at $1 \cdot 10^{-7}$~mbar to $2\cdot10^{-7}$~mbar). We then do a lift-off process in 80~$^{\circ}$C NMP for $>$~1 h and carefully rinse in ACE and IPA. 

\subsection{Scaffolding layer}
Now we pattern a scaffolding layer to fabricate the cross-overs. After prebaking, we spin LOR 5B at 3000~rpm and bake at 180~$^{\circ}$C for 5 min, followed by spinning PMMA 950k A2 at 4000~rpm and baking at 180~$^{\circ}$C for 5 min. We then beam write the negative pattern of the cross-over support structure using aligned electron beam lithography (200~nA beam, 25~nm fracturing, 1000~$\mu\mathrm{C}/\mathrm{cm}^2$ dose). The resist is developed using MIBK:IPA (1:3) for 1~min, and rinsed in IPA for 30~s. We then wet etch the scaffolding layer using MF-319 for 8~s, followed by a water rinse and IPA which stops the etch. Finally we remove the remaining PMMA layer with ACE (30~s) and reflow the LOR cross-over support layer on a hot plate at 200~$^{\circ}$C for 10~min. This creates a structurally stable arc shaped cross over scaffolding. 

\subsection{Coil wire patterning}
In order to pattern the narrow pitch coils, we spin PMMA 495 A8 at 2000~rpm, bake, spin PMMA 950k A2 at 2000 rpm and bake again. Then we lithographically define the coil wires, which overlap the capacitor wires (10~nA beam, 10~nm fracturing, 1800~$\mu\mathrm{C}/\mathrm{cm}^2$ dose) and develop the resist as described previously. Development is followed by the same plasma ashing, deposition of aluminum (1~nm/s, $120$~nm, p$\approx 2\cdot10^{-7}$~mbar) and lift off, during which the NMP (at 80~$^{\circ}$C, $3$ h) dissolves the LOR scaffolding layer.

\subsection{DC contact wire}
After a careful rinse with ACE and IPA we reproduce the previous layer recipe to pattern a small (500~nm$\times$4~$\mu$m) DC contact wire that symmetrically covers all overlap regions between capacitor wire and coil wire (two per coil and capacitor). Here we use an in-situ ion gun etch process (normal incidence with 4~cm gridded Kaufman source, 400~V, 21~mA for 5 min) right before the aluminum deposition of thickness 140~nm, in order to establish reliable contact. Contact is tested after lift-off on DC test structures of the same contact size located in the center of the chip. High resistance contacts with low capacitance at microwave frequencies would lead to additional parasitic in-series capacitances of the fabricated circuit. 

\subsection{Release}
For the final release step we prepare a silicon enriched solution of TMAH to selectively etch the silicon without aluminum corrosion \cite{GuiZhen2000,Fujitsuka2004}. We use a custom built reactor vessel with thermometer port and a hotplate with magnetic stir bar to mix 60~g of TMAH ($25$~\%, 6N) and 250~g water, and then add 5.12~g of silicon powder (-325 mesh, 5N) and stir at 300~rpm. After the chemical reaction calms down we start heating the solution up to 80~$^{\circ}$C. When the solution is clear, we wait for 1~h and prepare a clear mixture of 5.21~g of TMAH ($25$~\%, 6N) and 2.11~g of the oxidizing agent ammonium persulfate in a small beaker. We add the mixture to the solution (stir bar at 1000~rpm), wait 10~min to 15~min, reduce the stir speed and add the sample in a vertical position. The sample is securely clamped, but with the membranes open to a steady flow of solution on the back and front side of the chip. We keep the solution at 80~$^{\circ}$C and wait for the membranes to become fully transparent (1~h to 2~h). As a last step we carefully remove the sample, rinse it throughly in hot water, cold water, IPA, ultra purified IPA, and dry it using a CO$_2$ critical point dryer.

%%%%%%%%%%%%%%%%%%%%%%%%%

\section{Derivation of cavity response functions}
\label{app:D}

\subsection{Fourier Transform} 
We use the following convention for the Fourier transform. Given an operator $\hat{A}$ we define
\begin{align}
 \hat{A}(t)&=\frac{1}{\sqrt{2\pi}}\int_{-\infty}^{+\infty} d\omega e^{-i\omega t}\hat{A}(\omega)\\
\hat{A}(\omega)&=\frac{1}{\sqrt{2\pi}}\int_{-\infty}^{+\infty} dt e^{i\omega t}\hat{A}(t).
\end{align}

\subsection{Reflective coupling to a microwave resonator} \label{section:cavity}
\begin{figure}[b]
\begin{center}
\includegraphics[width=0.50\columnwidth]{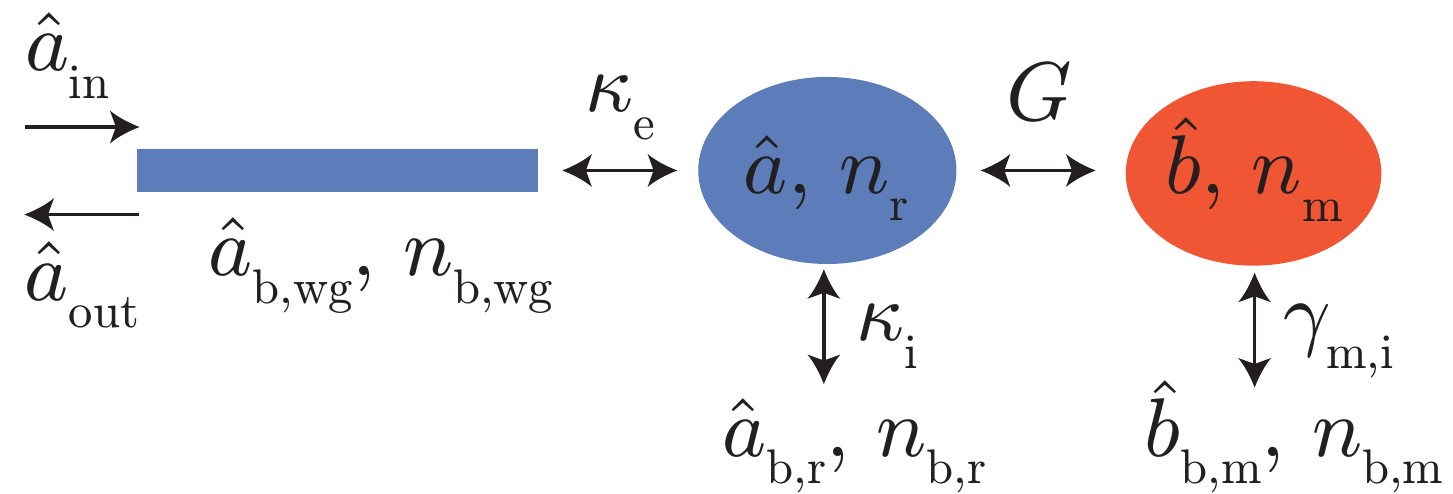}
\caption{\textbf{System modes, coupling rates and noise baths.} In the reflective geometry the microwave cavity mode $\aop$ is coupled to the coherent waveguide modes $\ain$ and $\aout$ with the external coupling strength $\kappaEe$. It is also coupled to a bath of noise photons, ideally at the refrigerator temperature $\nrb$, with the intrinsic coupling strength $\kappaEi$. In addition, the waveguide bath mode $\awb$ can be populated with thermal noise photons $\nwb$, which also couples with $\kappaEe$. The mechanical resonator mode $\bop$ is coupled to the microwave resonator with the parametrically enhanced electromechanical coupling strength $\GOM$. In addition, it is coupled to a bath of noise phonons, ideally at the refrigerator temperature $\nmb$, with the intrinsic coupling rate $\gammai$.
} 
\label{fig:schematic}
\end{center}
\end{figure}

We consider a resonator mode $\aop$ at frequency $\omegacE$, which is coupled to a single waveguide with coupling strength $\kappaEe$, and to the environment with the coupling strength $\kappaEi$ (see Fig.~\ref{fig:schematic}). We follow general input-output theory \cite{Walls1994} to write the time derivative of the annihilation operator
\begin{equation}
\dot{\aop}(t)  =  - \left ( i\omegacE + \frac{\kappa}{2} \right ) \hat{a}(t)  -\sqrt{\kappaEe} \ain(t) - \sqrt{\kappaEi} \ab(t) -\sqrt{\kappaEe} \awb(t),
\end{equation}
where $\kappa=\kappaEe+\kappaEi$ is the total resonator linewidth, $\ain(t)$ represents the annihilation operator of the coherent input mode, $\awb(t)$ the waveguide mode operator, and $\ab(t)$ is the respective field operator of the resonator environment. We take the Fourier transform to remove the time derivative, and simplify to get the frequency dependence
\begin{equation}
\aop(\omega)=\frac{-\sqrt{\kappaEe} \ain(\omega) -\sqrt{\kappaEi}\ab(\omega)-\sqrt{\kappaEe} \awb(t)}{\kappa/2+ i(\omegacE-\omega)}.
\label{eq:ares}
\end{equation}
The resonator output field is defined as 
\begin{equation}
\begin{aligned}
	\aout(\omega)
	&=\ain(\omega)+\sqrt{\kappaEe}\aop(\omega)\\
	&=\ain(\omega)+\frac{-\kappaEe \ain(\omega) -\sqrt{\kappaEe \kappaEi }\ab(\omega)-\kappaEe \awb(t)}{\kappa/2+ i(\omegacE-\omega)},
\end{aligned}
\end{equation}
which we can use to calculate the complex scattering parameter as measured by a network analyzer
\begin{equation}
S_{11}(\omega)=\frac{\left<\aout(\omega)\right>}{\left<\ain(\omega)\right>}=1-\frac{\kappaEe}{\kappa/2+ i(\omegacE-\omega)},
\label{eq:barecavity}
\end{equation}
where the incoherent bath mode terms drop out. We use this function to simultaneously fit the real and imaginary part of the measured cavity response and extract the intrinsic and extrinsic cavity coupling rates.

\subsection{Drive photon number}
It is useful to define the intra-cavity photon number $\ncavd$ due to a classical coherent drive tone at frequency $\omegad$. We replace the field operators in Eq.~\ref{eq:ares} with the classical amplitudes $\hat{a}(\omega)\rightarrow \alpha(\omega)$ and discard the resonator and waveguide bath modes to get 
\begin{equation}
\ncavd=|\alpha_\mathrm{d}|^2=|\alpha_\mathrm{in}|^2\frac{4\kappaEe}{\kappaE^2+4\Deltard^2}.
\label{eq:drivephotons}
\end{equation}
Here we have introduced the resonator drive detuning $\Deltard=\omegacE-\omegad$ and the input photon flux $|\alpha_\mathrm{in}|^2= P_\mathrm{in}/(\hbar \omegad)$. The power at the cavity input can be expressed as $P_\mathrm{in}=10^{-3} 10^{(\mathcal{A} + P_\mathrm{d})/10}$ with $P_\mathrm{d}$ the drive power in dBm and $\mathcal{A}$ the total attenuation of the input line in dB.

\subsection{Asymmetric lineshape}
Fano line shapes generally originate from interference between a resonant mode and a background mode \cite{Fano1866}. Experimental imperfections, such as leakage or reflections in the feedline circuit, can lead to such asymmetric cavity line shapes. We can model this effect by introducing a complex valued external resonator to waveguide coupling parameter $\bar{\kappa}_\mathrm{e}=|\kappaEe|e^{- i q}$, where $q$ is a version of the Fano parameter. While small $q$ values do not change the magnitude of the inferred external coupling (or the drive photon number), they correctly model small asymmetries in the Lorentzian cavity response \cite{Geerlings2012}. For simplicity we define the generalized coupling $\bar{\kappa}_\mathrm{e}=\kappaEe-i q$ and substitute into Eq.~\ref{eq:barecavity}, to get the generalized resonator line shape
\begin{equation}
S_{11}(\omega)=1-\frac{\kappaEe - i q }{\kappa/2- i(\omegacE-\omega)}.
\end{equation}

%%%%%   
\section{Derivation of cavity electromechanical response functions}
\label{app:E}

In this section we follow previous work \cite{Marquardt2007,Dobrindt2008,Teufel2011,Rocheleau2009} to calculate the coherent response and the full noise spectrum of the system. In contrast to earlier treatments we also include thermal noise in the feedline circuit, which gives rise to an increased mechanical occupation and an asymmetric cavity noise line shape. 

\subsection{System Hamiltonian and equations of motion}
The Hamiltonian of the coupled microwave cavity-mechanical system (see Fig.~\ref{fig:schematic}) can be written as 
\begin{equation}
\hat{H}=\hbar \omegacE \hat{a}^{\dagger}\hat{a}+\hbar \omegam \hat{b}^{\dagger}\hat{b} + \hbar \gzeroE \hat{a}^{\dagger}\hat{a}(\hat{b}^{\dagger}+\hat{b}),
\end{equation}
where $\hat{b}$ ($\hat{b}^\dagger$) is the annihilation (creation) operator of the mechanical mode at frequency $\omegam$, and $\gzeroE$ is the electromechanical coupling strength, i.e.~the resonator frequency shift due to a mechanical displacement corresponding to half a phonon on average. We excite the microwave resonator mode using a strong drive tone at frequency $\omega_\mathrm{d}$, detuned from the resonator frequency by $\Deltard=\omegacE-\omegad$. The linearized Hamiltonian in the rotating frame is then given as 
\begin{equation}
\hat{H'}= - \hbar \Deltard \hat{a}^{\dagger}\hat{a}+\hbar \omegam \hat{b}^{\dagger}\hat{b} + \hbar \GOM(\hat{a}^{\dagger}+\hat{a})(\hat{b}^{\dagger}+\hat{b}),
\end{equation}
where $\GOM=\sqrt{n_\mathrm{d}}\gzeroE$ is the parametrically enhanced optomechanical coupling strength.
The linearized Langevin equations are given as
\begin{align}
\dot{\hat{a}}(t)  & =  - \left ( i \Deltard + \frac{\kappa}{2} \right ) \hat{a}(t) - i \GOM (\hat{b}(t) + \hat{b}^{\dagger}(t))-\sqrt{\kappaEe} \ain(t) - \sqrt{\kappaEi} \ab(t) -\sqrt{\kappaEe} \awb(t)\\
\dot{\hat{b}}(t) & = - \left ( i \omegam + \frac{\gammai}{2} \right ) - i \GOM (\hat{a}^{\dagger}(t)+ \hat{a}(t) )- \sqrt{\gammai} \bb(t).
\end{align}
Taking the Fourier transform and simplifying we obtain
\begin{align}
\chir^{-1}(\omega){\hat{a}(\omega)}  & =  - i \GOM (\hat{b}(\omega) + \hat{b}^{\dagger}(\omega))-\sqrt{\kappaEe} \ain(\omega) - \sqrt{\kappa_\mathrm{i}} \ab(\omega) \\
\chirt^{-1 }(\omega){\hat{a}^{\dagger}(\omega)}  & =   i \GOM (\hat{b}(\omega) + \hat{b}^{\dagger}(\omega))-\sqrt{\kappaEe} \ain^{\dagger} (\omega)- \sqrt{\kappai} \ab^{\dagger}(\omega)\\
\chim^{-1 }(\omega){\hat{b}(\omega)} &  =  - i \GOM (\hat{a}(\omega) + \hat{a}^{\dagger}(\omega))-\sqrt{\gammai} \bb(\omega)\\
\chimt^{-1 }(\omega){\hat{b}^{\dagger}(\omega)} & =    i \GOM (\hat{a}(\omega) + \hat{a}^{\dagger}(\omega))-\sqrt{\gammai} \bb^{\dagger}(\omega),
\end{align}
where we have introduced the uncoupled susceptibilities of the cavity and the mechanical mode	\begin{align}
\chir^{-1}(\omega) & =\kappa/2 + i ( \Deltard-\omega)\\
\chirt^{-1}(\omega) & =\kappa/2 - i (\Deltard+\omega)\\
\chim^{-1}(\omega) & =\gammai/2 + i (\omegam- \omega)\\
\chimt^{-1}(\omega) & =\gammai/2 - i (\omegam+\omega).
\end{align} 
In the sideband resolved limit $\omegam\gg\kappaE, \GOM$, and for positive detuning of the drive tone $\Deltard\approx \omegam$ (red side pumping), the linearized Langevin equations can be written approximately as,
\begin{align}
\hat{a}(\omega)&=\frac{i \GOM \chim \chir \sqrt{\gammai} \bb(\omega) - \chir (\sqrt{\kappae} \ain(\omega) + \sqrt{\kappai}\ab(\omega)+\sqrt{\kappae} \awb(\omega))} {1+\GOM^2 \chim \chir}
\label{eq:linLang1}\\
\hat{b}(\omega)&=\frac{- \chim \sqrt{\gammai} \bb(\omega) - i \GOM \chim \chir (\sqrt{\kappae}\ain(\omega)+\sqrt{\kappai} \ab(\omega)+\sqrt{\kappae} \awb(\omega)) }{1+\GOM^2 \chim \chir}.
\label{eq:linLang2}
\end{align}
Now we can calculate the cavity output mode 
\begin{equation}
\begin{aligned}
\hat{a}_\mathrm{out}(\omega)&=\ain(\omega) + \sqrt{\kappae} \hat{a}(\omega)\\
&=\ain(\omega)-(\ain(\omega)+\awb(\omega) )\frac{\kappae \chir}{1+\GOM^2 \chim \chir}
- \ab(\omega) \frac{\sqrt{\kappae\kappai} \chir}{1+\GOM^2 \chim \chir}
+ \bb(\omega) \frac{i \GOM \sqrt{\kappae \gammai} \chim \chir}{1+\GOM^2 \chim \chir}.
\end{aligned}\label{eq:aout}
\end{equation}
%%%%%%%%%%%%%%%%%%%%%%%%%%%%
	
\subsection{Electromagnetically Induced Transparency}
We first calculate the coherent part of the system response using Eq.~\ref{eq:aout} and drop incoherent noise terms to get
\begin{equation}
S_{11}(\omega)=\frac{\left<\aout(\omega)\right>}{\left<\ain(\omega)\right>}=
1-\frac{\kappae\chir}{1+\GOM^2 \chim \chir}.
\end{equation}
Substituting the bare response of the cavity and the mechanical oscillator we get the coherent EIT response function valid for small probe drive strengths
\begin{equation}
S_{11}(\omega)=1-\frac{\kappae}{\kappa/2+i(\Deltard-\omega)+\frac{\GOM^2}{\gammai/2+i(\omegam-\omega)}}.
\end{equation}
In order to take into account potential interference with a continuum of parasitic modes, we can follow the procedure outlined above. Substituting $\kappae\rightarrow\kappae-iq$ we get 
\begin{equation}
S_\mathrm{11, as}(\omega)=1-\frac{\kappae- i q}{\kappa/2+i(\Deltard-\omega)+\frac{\GOM^2}{\gammai/2+i(\omegam-\omega)}},
\end{equation}
which can be used to fit asymmetric EIT spectra. 
%For the results presented in the main paper this was not necessary and we set $q\rightarrow 0$.
%%%%%%%%%%%%%%%%%%%%%%%%%%%%%

\section{Quantum derivation of observed noise spectra}
\label{app:F}
Using the Fourier transforms defined above, we can write the spectral density of an operator $\hat{A}$ as
\begin{align}
S_{AA}(t)&=\int_{-\infty}^{+\infty} d\tau e^{i\omega\tau}\left<\hat{A}^{\dagger}(t+\tau)\hat{A}(t)\right>\label{Eq:TimePSD}\\
S_{AA}(\omega)&=\int_{-\infty}^{+\infty}d\omega'\left<\hat{A}^{\dagger}(\omega)\hat{A}(\omega')\right>.
\end{align}
The auto-correlation function of the detected normalized field amplitude (or the photo current) of the output mode $\hat{I}(t)=\aout(t)+\aout^{\dagger}(t)$ is then given as
\begin{equation}
S_{II}=\int_{-\infty}^{+\infty} d\omega'\left<\left(\aout(\omega)+\aout^{\dagger}(\omega)\right)\left(\aout(\omega')+\aout^{\dagger}(\omega')\right)\right>.
\label{Eq:SII}
\end{equation}
Substituting $\aout(\omega)$ and $\aout^{\dagger}(\omega)$ from Eq.~\ref{eq:aout} we find a general expression for the single sided noise spectrum
\begin{multline}
S(\omega)=
\nwb \Big|\Big(1-\frac{\kappae \chirt}{1+\GOM^2 \chimt \chirt}\Big)\Big|^2 
+ \nrb \frac{|\kappae \kappai \chirt|^2 }{|1+\GOM^2 \chimt \chirt|^2}
+ \nmb \frac{\kappae \gammai \GOM^2 |\chimt|^2 |\chirt|^2}{|1+\GOM^2 \chimt \chirt|^2} \\
+(\nwb+1)  \Big|\Big(1-\frac{\kappae \chir}{1+\GOM^2 \chim \chir}\Big)\Big|^2 
+(\nrb+1) \frac{\kappae \kappai |\chir|^2 }{|1+\GOM^2 \chim \chir|^2} 
+ (\nmb+1) \frac{\kappae \gammai \GOM^2 |\chim|^2 |\chir|^2}{|1+\GOM^2 \chim \chir|^2}.
\label{Eq:FullS}
\end{multline}
Here, $\nwb$ and $\nrb$ represent the bath of noise photons from the waveguide and the microwave resonator environment respectively; $\nmb$ corresponds to the phonon bath at the mechanical frequency  (see Fig.~\ref{fig:schematic}). We assume thermal input noise correlations for all input noise terms, 
i.e.~$\langle \bb(\omega)\bb^\dagger(\omega')\rangle=(\nmb+1)\delta(\omega+\omega')$, 
$\langle \bb^\dagger(\omega)\bb(\omega')\rangle=\nmb\delta(\omega+\omega')$,
$\langle \ab(\omega)\ab^\dagger(\omega')\rangle=(\nrb+1)\delta(\omega+\omega')$,
$\langle \ab^\dagger(\omega)\ab(\omega')\rangle=\nrb\delta(\omega+\omega')$, 
$\langle \awb(\omega)\awb^\dagger(\omega')\rangle=(\nwb+1)\delta(\omega+\omega')$ and 
$\langle \awb^\dagger(\omega)\awb(\omega')\rangle=\nwb\delta(\omega+\omega')$.

In the sideband resolved regime and positive detuning (red sideband pump) we can drop the terms proportional to $\chimt$ and $\chirt$. In order to represent a realistic experimental setup, we introduce the fixed gain $\mathcal{G}$ in units of dB and the system noise temperature $n_\mathrm{add}$ in units of resonator quanta and referenced to the cavity output. We can now write the full expression for the single sided power spectral density as measured by a spectrum analyzer, valid in the presence of all relevant noise baths
\begin{multline}
S(\omega)=
\hbar \omegad 10^{\mathcal{G}/10} 
\Big[
n_\mathrm{add} + \nwb 
+(\nwb+1)  \Big|\Big(1-\frac{\kappaEe \chir}{1+\GOM^2 \chim \chir}\Big)\Big|^2\\
+(\nrb+1) \frac{\kappaEe \kappai |\chir|^2 }{|1+\GOM^2 \chim \chir|^2}
+(\nmb+1)\frac{\kappaEe \gammai \GOM^2 |\chim|^2 |\chir|^2}{|1+\GOM^2 \chim \chir|^2}
\Big].
\label{Eq:FullS2}
\end{multline}
We minimize the number of free parameters by eliminating the resonator bath $\nrb$ using the relation
\begin{equation}
\kappa n_\mathrm{r}=\kappae \nwb + \kappai \nrb.
\label{eq:baths}
\end{equation}
With the Eqs.~\ref{eq:linLang1} and \ref{eq:linLang2} we can calculate \cite{Dobrindt2008} the mechanical occupation $\nbar$
\begin{equation}
\nbar=\nmb \left(\frac{\gammai}{\kappaE}\frac{4\GOM^2+\kappaE^2}{4\GOM^2+\kappaE\gammai}\right) + \nr\left( \frac{4\GOM^2}{4\GOM^2 +\kappaE\gammai}\right),
\label{eq:bath}
\end{equation}
which we use to also replace the mechanical bath occupation $\nmb$ in Eqs.~\ref{Eq:FullS2}. 
%This is the model we use for fitting the experimental data at high drive powers.

\subsection{Thermal waveguide noise}
At large drive photon numbers we observe a power dependent increase of the measured noise background. Using a cavity filter to remove broad band phase noise of the microwave source did not remove this background. Similarly, the power levels of the observed microwave signals are believed to be far from saturating the HEMT amplifier. We therefore take the most conservative approach and attribute this background rise entirely to a power dependent thermal waveguide photon bath $\nwb$. Such a noise term could originate from a rise of the electronic temperature of the on-chip feedline circuit.

\begin{figure}[h]
\begin{center}
\includegraphics[width=1.0 \columnwidth]{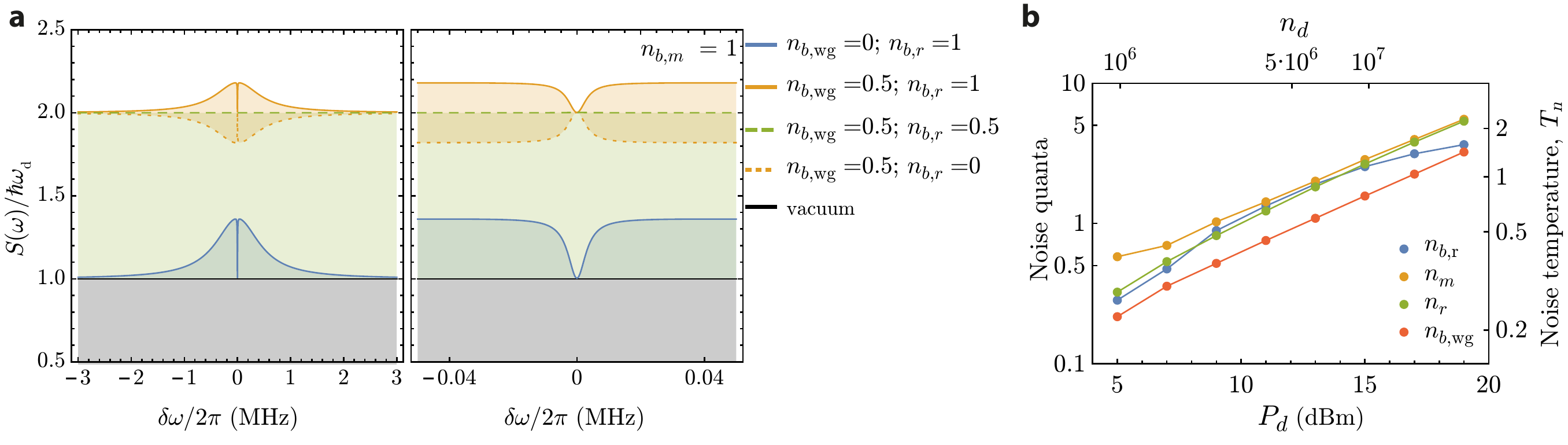}
\caption{\textbf{Noise Budget. a,}~Shown are solutions to Eq.~\ref{Eq:FullS2} over the resonator bandwidth (left) and the mechanical bandwidth (right). The different curves are plotted for a set of different noise bath parameters, as indicated in the legend. \textbf{b,}~The fitted noise occupancies $\nbar$, $\nr$ and $\nwb$ are shown together with the resonator bath $\nrb$.} 
\label{fig:noise}
\end{center}
\end{figure}

Figure~\ref{fig:noise}~a shows a comparison of noise spectra with and without waveguide noise. In the presence of a broad band thermal input field, the background rises and the resonator noise peak shrinks, even though the resonator noise bath is kept constant. The reason is that the cavity filters the broad band input noise, effectively changing the background of the cavity noise peak. If the waveguide noise bath matches the cavity noise bath, no cavity noise peak is observed even though the cavity temperature is increased.

Compared to a model which attributes this background change to a modification of the amplifier noise temperature or an increased attenuation at the output of the detection circuit, i.e.~a change of $\nadd$ only, we extract almost twice the cavity occupation $\nr$ using Eq.~\ref{Eq:FullS2}. This also affects the lowest observed mechanical occupation and raises it from $\nbar = 0.33$ to 0.58. Figure~\ref{fig:noise}~b shows the extracted noise baths for the dataset measured at 26 mK. 

\subsection{Asymmetric noise spectra}
In our measurements the cavity noise bath exceeds the waveguide noise bath for all relevant pump powers. In this case the microwave resonator bath $\nr$ manifests itself as a broad band resonator noise peak on top of the background. This power dependent noise peak shows a small asymmetry for the highest pump powers. Such an asymmetry is qualitatively expected from interference between narrow band cavity noise and broad band waveguide noise. We follow a similar procedure as outlined above and introduce a complex waveguide coupling constant to find better agreement with the measured data in this limit. We make the substitution $\kappae\rightarrow\kappae-iq$ in the first term proportional to $\nwb$ in Eq.~\ref{Eq:FullS2} and expand it. For $\kappaEe^2\gg q^2$ we can only keep $q$ to linear order and simplify the expressions. We can then write the asymmetric noise power spectrum with two additional terms as
\begin{equation}
S_\mathrm{as}(\omega)=S(\omega) 
+ \hbar \omegad 10^{\mathcal{G}/10} 
(\nwb+1)\Big[
\frac{2 q \GOM^2 (\omegam-\omega) |\chim|^2 |\chir|^2}{|1+\GOM^2 \chim \chir|^2}
+\frac{2 q (\Deltard-\omega) |\chir|^2}{|1+\GOM^2 \chim \chir|^2}
\Big].
\end{equation}
The additional two terms are odd functions with a vanishing integral. This ensures the same fit results as obtained compared to using the symmetric model  Eq.~\ref{Eq:FullS2}. Defined in this way, the asymmetry scales with the waveguide noise bath and the Fano parameter $q$, which is independent of any other parameters. We find excellent agreement between this model and the measured broad band noise spectra at high pump powers (see main text). It is important to point out that only the relevant bath occupancies $\nbar$, $\nr$, $\nwb$ as well as $q$ (in the case of the highest drive powers) are actual fit parameters, while all other relevant parameters are extracted from - or verified in - independent (low drive power) measurements. 

\subsection{Relation to the displacement spectrum}
In the weak coupling regime we can relate the single sided displacement spectrum $S_\mathrm{x}(\omega)$ using
\begin{equation}
\frac{S(\omega)}{\hbar \omega} =  \frac{S_\mathrm{x}(\omega)}{\xzpf^2} \frac{\kappaEe}{\kappaE} \Gamma_\mathrm{+}
\end{equation}
with the photon scattering rate $\Gamma_\mathrm{+}\approx 4\GOM^2/\kappaE$ for the optimal detuning $\Deltard=\omegam$, and the factor $\kappaEe/\kappaE$ taking into account the limited collection efficiency of photons leaving the cavity.

\section{Low drive power limits}
\label{app:G}
At low drive powers and sufficient shielding from room temperature Johnson noise, it is a very good approximation to set $\nwb \rightarrow 0$. Eliminating the waveguide noise input allows for a significant simplification of the power spectrum
\begin{equation}
S(\omega) = 
\hbar \omegad 10^{\mathcal{G}/10} 
\Big(
1
+n_\mathrm{add}
+\frac{4 \kappae (\nr \kappa (\gammai^2 + 4 (\omegam - \omega)^2) 
+4\nmb\gammai \GOM^2)}
{|4 \GOM^2 + (\kappa + 2 i (\Deltard - \omega))(\gammam + 2 i (\omegam - \omega))|^2}
\Big),
\label{Eq:fullSsimplified}
\end{equation}
with the mechanical noise bath $\nmb$ related to the mechanical occupation $\nbar$ in Eq.~\ref{eq:bath}. At low drive photon numbers we see no indication of TLS, pump phase noise, or waveguide heating of the cavity. The chosen attenuation and shielding of input and output microwave lines connecting the sample to room temperature Johnson noise limits the expected cavity occupation to $\nr \ll 0.05$ (see for example Ref.~\cite{Fink2010} for an independent temperature measurement in a similar setup with less attenuation). Under these assumptions, which are verified by our low power measurements (constant background noise, no cavity noise peak), we can simplify the power spectrum to the more standard form used in cavity electro- and optomechnics 
\begin{equation}
S(\omega) = 
\hbar \omegad 10^{\mathcal{G}/10} 
\Big(
1
+n_\mathrm{add}
+\frac{16 \nmb \kappae \gammai \GOM^2}
{|4 \GOM^2 + (\kappa + 2 i (\Deltard - \omega))(\gammam + 2 i (\omegam - \omega))|^2}
\Big).
\label{Eq:fullSsimplified2}
\end{equation}
Without resonator occupation, Eq.~\ref{eq:bath} simplifies to
\begin{equation}
\nbar=\nmb \left(\frac{\gammai}{\kappaE}\frac{4\GOM^2+\kappaE^2}{4\GOM^2+\kappaE\gammai}\right) 
\approx \nmb \Big(\frac{1}{C+1}\Big),
\label{eq:bath2}
\end{equation}
where we introduced the cooperativity $C=4\GOM^2/(\kappaE \gammai)$ and assumed moderate coupling strength $4\GOM^2 \ll \kappaE^2$ in the last step.

\section{Linear response limit - system calibration}
\label{app:H}
For very small drive powers  where $C\ll1$ we can simplify the expected thermal noise spectrum further. Dropping terms associated with backaction allows to measure the displacement noise in a self-calibrated way. This compact model is particularly useful to back out $\gzeroE$ and the system noise temperature with a minimal number of assumptions. 

Starting with Eq.~\ref{Eq:fullSsimplified2} we can make the replacement $\nmb \approx \nbar$ and drop the backaction term in the denominator. Both is valid for $C\rightarrow0$. We then insert $\GOM=\sqrt{\ncavd}\gzeroE$ with the drive photon number defined in Eq.~\ref{eq:drivephotons}. We furthermore assume that the gain of the system is flat over the relevant detuning such that we can introduce the directly reflected pump power measured at the spectrum analyzer $P_\mathrm{r}=10^{\mathcal{G}/10} P_\mathrm{out}$. The cavity output power is related to the cavity input power via Eq.~\ref{eq:barecavity} 
\begin{equation}
|S_{11}|^2=\frac{P_\mathrm{out}}{P_\mathrm{in}}=\frac{4\Deltard^2+(\kappaE-2\kappaEe)^2}{4\Deltard^2+\kappa^2}.
\end{equation}
Finally, we can write the measured noise spectrum, normalized by the measured reflected pump tone 
\begin{equation}
\frac{S(\omega)}{P_\mathrm{r}}= 
\mathcal{O}+
\frac{64 \nbar \kappaEe^2 \gammai \gzeroE^2}
{
(4\Deltard^2 +(\kappaE-2\kappaEe)^2)       
(\kappaE^2+4(\Deltard-\omega)^2)
(\gammai^2+4(\omegam-\omega)^2}.
\label{eq:calibration}
\end{equation}
Only directly measurable system parameters and the temperature of the mechanical mode need to be known to extract $\gzeroE$ without any further assumptions about the particular gain, attenuation or noise temperature of the chosen measurement setup. Knowing $\gzeroE$, it is easy to accurately back out the input attenuation $\mathcal{A} = -66.3$~dB and drive photon number $\ncavd$ (for example from an EIT measurement). Furthermore, from the measured offset 
\begin{equation}
\mathcal{O}=(1+n_\mathrm{add}) \frac{4\kappaEe}{\ncavd (4\Deltard^2 + (\kappaE-2\kappaEe)^2)}
\end{equation} 
we conveniently infer the system noise temperature in units of photons $n_\mathrm{add}\approx 30$. The system gain $\mathcal{G}\approx 46$~dB is then easily determined from the not-normalized wide band background of the measured noise spectrum $S(\omega)$. 

%%%%%%%%%%%%%%%%%%%%%%%%%%%%%%%%%%%%%

\section{Vacuum Rabi splitting and ac Stark tuning of a nanoscopic two level system}
\label{app:I}
The demonstrated motional sideband cooling is facilitated by the small capacitor gap size, the high impedance and small stray capacitance of the electrical circuit. These are very desirable properties in the context of quantum electrodynamics (QED), e.g.~with atomic systems. The vacuum fluctuations of the tested resonator give rise to a large root mean square voltage of $V_\mathrm{vac}=\sqrt{\hbar\omegacE/(2C_\mathrm{tot})}=\omegacE\sqrt{\hbar Z_\mathrm{tot}/2}\approx 21$~$\mu$V. With a gap size of only $s\approx 80$~nm, the electric field across the capacitor $C_\mathrm{m}$ is as large as $E_\mathrm{vac}\approx 260$~V/m, about $10^3$ times larger than in typical coplanar waveguide resonators and $ >10^5$ times that of typical small 3D microwave cavities of similar frequency. For the largest pump photon numbers $\ncavd\approx 4\cdot10^7$ we apply an rms voltage of $V_\mathrm{max}\approx 1.7$~kV, corresponding to a maximum field strength of $E_\mathrm{max} \approx  21$~GV/m. To our knowledge this is a record high value for a low loss superconducting microwave resonator.

\begin{figure*}[b]
\begin{center}
\includegraphics[width=0.6\columnwidth]{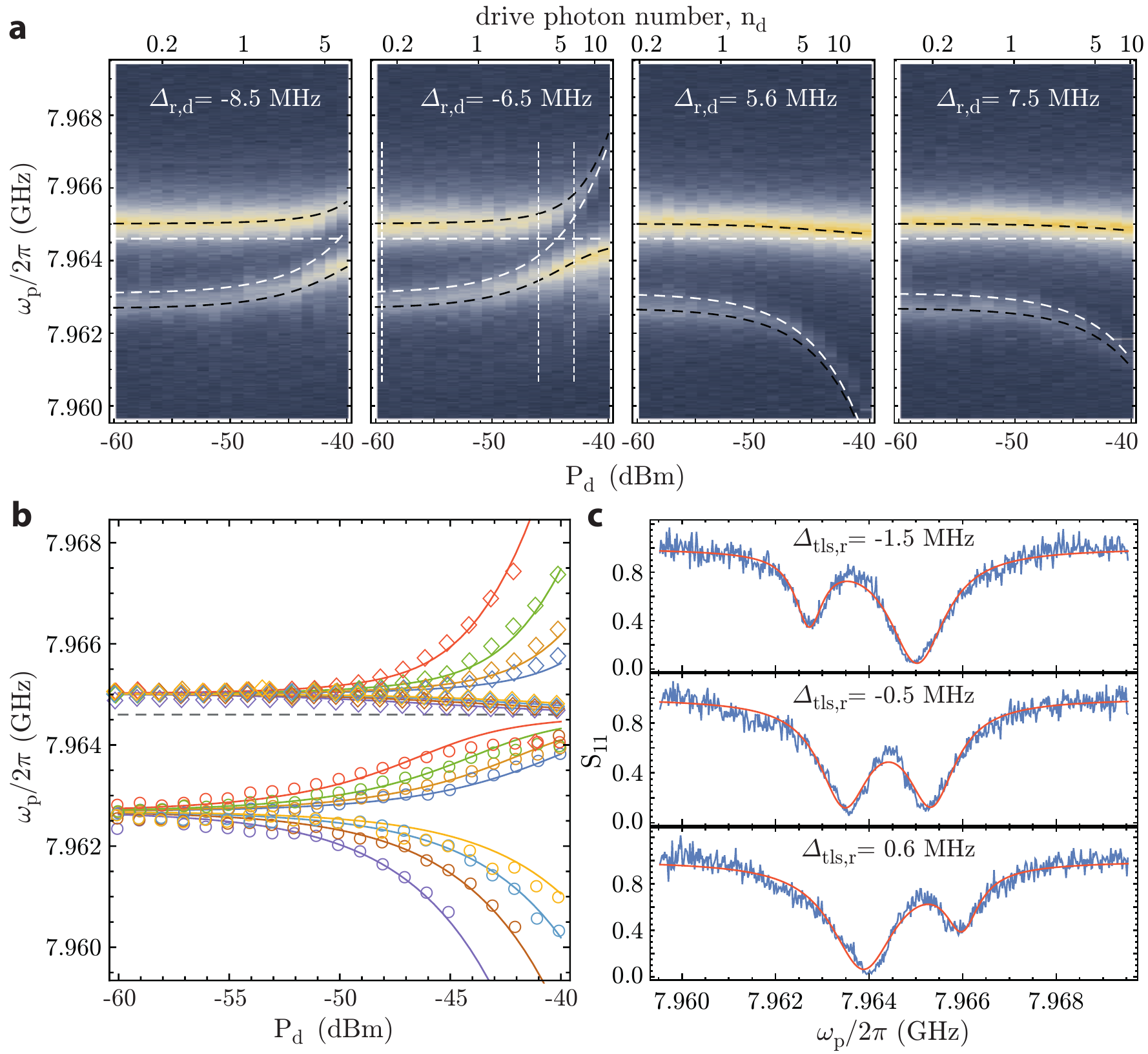}
\caption{\textbf{Two-level-system vacuum Rabi mode splitting a,} Color plot of the measured reflection $S_{11}$ (blue is high, yellow is low) of a weak coherent probe tone $\omega_\mathrm{p}$  as a function of the drive power $P_\mathrm{d}$ for 4 different drive detunings $\Deltard$. Dashed lines show uncoupled (white) and coupled TLS-resonator eigenvalues. \textbf{b,} Measured eigenfrequencies extracted from a double Lorentzian fit (circles), theory (solid lines) and the bare resonator frequency (dashed line) for different negative $\Deltard=(-8.5, -7.5, -6.5, -5.5)$~MHz and positive $\Deltard=(4.5, 5.5, 6.5, 7.5)$~MHz drive detunings. 
\textbf{c,} Measured spectrum (blue) and double Lorentzian fit (red) for $\Deltard=-6.5$~MHz and three different drive powers, see vertical dotted lines in panel a.} 
\label{fig:tls}
\end{center}
\end{figure*}

Large vacuum fields enable efficient dipole coupling $\gzeroTLS = \vec{E} \cdot \dipoleTLS$ to microscopic systems with small dipole moments $\dipoleTLS$, such as molecular two level systems (TLS) \cite{Sarabi2015b}, single atoms, or charge quantum dots. We demonstrate this by observing coherent coupling between the resonator vacuum field and a nanoscopic TLS on the substrate (see Fig.~\ref{fig:tls}). The demonstrated ac-Stark control of this nanoscopic system complements previous strain based control techniques \cite{Grabovskij2012}, and will be useful to suppress the negative impact of TLS in engineered solid state quantum systems. The quantum nonlinearity of the TLS furthermore represents a unique calibration tool, which is generally missing in electro- and opto-mechanical systems. More importantly, it could point the way how to better understand the origin and location of TLS, which are believed to be the limiting factor not only for ground state cooling of electromechanical systems but more generally for improving the integration density of low loss superconducting circuits.

In order to observe vacuum Rabi mode splitting, we reduce the probe power significantly such that the probe photon number $\ncavp \ll 1$. In order to control the interaction, we use a second off-resonant drive tone at frequency $\omegad$ to Stark-shift the TLS into and out of resonance with the resonator. For sufficient detuning $\DeltaTLSd \equiv \omegaTLS-\omegad$ we can adiabatically eliminate direct transitions of the TLS \cite{Blais2007} and define the new Stark-shifted TLS frequency as
\begin{equation}
\omegatildeTLS \approx \omegaTLS + \frac{\OmegaR^2}{2\DeltaTLSd},
\label{eq:TLS}
\end{equation} 
\noindent with $\OmegaR= 2\gzeroTLS \sqrt{\ncavd}$ the Rabi frequency due to the off-resonant drive tone. 

Using this linearized model we find very good agreement for different drive detunings $\Deltard \equiv \omegacE-\omegad$ and drive strength in the range $|\OmegaR/(2\DeltaTLSd)| < 1$ for which the linearization is valid, see Fig.~\ref{fig:tls}. The attenuation $\mathcal{A}=-66.8$~dB, entering via the drive photon number $\ncavd$, is the only fit parameter and agrees with our previous calibration to within 0.5~dB. The presented Stark-shift measurements therefore independently confirm the previously calibrated drive photon numbers and the electromechanical coupling $\gzeroE$.  Similar Stark-shift calibrations are quite common in superconducting qubit experiments \cite{Schuster2005}. Opto- and electromechanics experiments on the other hand typically lack the necessary strong vacuum nonlinearities for an absolute calibration with the vacuum field. 

It is now clear that the presence of a coupled TLS can strongly modify the resonator lineshape. This could explain why at low drive powers, where the TLS happens to be close to the resonator frequency, the electromechanical transduction efficiency is reduced and the mechanical occupancy appears to be lower than expected (see Fig.~4~c in the main text). Although the studied TLS should be far detuned in the relevant power range, there may be other weakly coupled TLS which do not Stark tune as easily, but still  absorb and scatter photons at a high rate. At high drive powers on the other hand, all TLS are either saturated due to the high intra-cavity photon numbers, or far detuned from the drive and resonator frequencies. 

On resonance, the vacuum Rabi split linewidths are approximately given by an equal mixture of resonator and TLS linewidth, which we use to estimate $\gammaTLS/(2\pi)=1.3$~MHz. We extract a TLS coupling of $\gzeroTLS/(2\pi) = 0.9$~MHz from the minimum separation of the measured spectroscopic lines (see Fig.~\ref{fig:tls}~c).  This puts our system very close to the strong coupling limit $\gzeroTLS \geq (\kappaE,\gammaTLS)$. We can also put a lower bound on the electric dipole moment of this TLS $|d| \geq 0.7$~Debye depending on the TLS orientation and the exact position on the capacitor. 

This value is comparable to many atomic, nano- and microscopic systems, where strong coupling is not readily observed. The presented techniques therefore open up new possibilities to realize hybrid systems and can help study the loss mechanism that plague superconducting quantum processors more directly, selectively and with higher sensitivity.
%%%%%%%%%%%%%%%%%%%%%%%%%%%%%

\end{document}